\pdfoutput=1
\begingroup\expandafter\expandafter\expandafter\endgroup
\expandafter\ifx\csname pdfsuppresswarningpagegroup\endcsname\relax
\else
\fi

\documentclass[preprint]{elsarticle} %  preprint, review, 11pt

\usepackage[utf8]{inputenc}
\usepackage[T1]{fontenc}
\usepackage[english]{babel}
\usepackage{csquotes}
\usepackage{textcomp}

\usepackage[usenames,dvipsnames]{xcolor}

\usepackage[hyphens]{url}
\biboptions{sort&compress, square, comma}
\bibpunct{{\unskip~[}}{]}{,}{n}{}{;} % put the ~ in "~[1]" so you don't have to. First removes a space if there is one there.
\usepackage{threeparttable}

\usepackage[breaklinks=true, linkcolor=blue, citecolor=blue, colorlinks=true]{hyperref}
\usepackage{graphicx}
\usepackage{caption,subcaption}
\usepackage[version=4]{mhchem}
\usepackage{booktabs,multirow}
\usepackage{latexsym,amsmath,amssymb}

\usepackage[textsize=small]{todonotes}  % use option [disable] to hide them

\usepackage[retainorgcmds]{IEEEtrantools}

\usepackage[binary-units]{siunitx}
\DeclareSIUnit\atm{atm}
\DeclareSIUnit\bar{bar}
\DeclareSIUnit{\calorie}{cal}

\usepackage{algorithm2e}
\usepackage{scrextend}
\usepackage{tikz}
\usetikzlibrary{decorations.pathmorphing}
\tikzset{snake it/.style={decorate, decoration=snake}}
\usepackage{environ}
\makeatletter
\newsavebox{\measure@tikzpicture}
\NewEnviron{scaletikzpicturetowidth}[1]{%
  \def\tikz@width{#1}%
  \begin{lrbox}{\measure@tikzpicture}%
  \BODY
  \end{lrbox}%
  \pgfmathparse{#1/\wd\measure@tikzpicture}%
  \BODY
}
\makeatother

\graphicspath{ {Figures/} }

\usepackage{array}

\usepackage{bm}
\newcommand{\dma}{\bm{D}^{\text{MA}}}
\newcommand{\dmc}{\bm{D}^{\text{MC}}}

\usepackage[normalem]{ulem}

\journal{Journal of Computational Physics}
\begin{document}

\begin{frontmatter}

\title{A fast, low-memory, and stable algorithm for implementing multicomponent transport in direct numerical simulations}

\author[osu]{Aaron J.~Fillo}
\author[galcit]{Jason Schlup}
\author[caltech]{Guillaume Beardsell}
\author[caltech]{Guillaume Blanquart}
\author[osu]{Kyle E.~Niemeyer\corref{cor1}}
\ead{kyle.niemeyer@oregonstate.edu}

\address[osu]{School of Mechanical, Industrial, and Manufacturing Engineering, Oregon State University, Corvallis, OR 97331, USA}
\address[galcit]{Graduate Aerospace Laboratories, California Institute of Technology, Pasadena, CA USA}
\address[caltech]{Department of Mechanical Engineering, California Institute of Technology, Pasadena, CA, USA}

\cortext[cor1]{Corresponding author}

%====================================================================
\begin{abstract}
Implementing multicomponent diffusion models in reacting-flow simulations is computationally expensive due to the challenges involved in calculating diffusion coefficients.
Instead, mixture-averaged diffusion treatments are typically used to avoid these costs.
However, to our knowledge, the accuracy and appropriateness of the mixture-averaged diffusion models has not been verified for three-dimensional turbulent premixed flames.
In this study we propose a fast, efficient, low-memory algorithm and use that to evaluate the role of multicomponent mass diffusion in reacting-flow simulations.
Direct numerical simulation of these flames is performed by implementing the Stefan--Maxwell equations in NGA.
A semi-implicit algorithm decreases the computational expense of inverting the full multicomponent ordinary diffusion array while maintaining accuracy and fidelity.
We first verify the method by performing one-dimensional simulations of premixed hydrogen flames and compare with matching cases in Cantera. 
We demonstrate the algorithm to be stable, and its performance scales approximately with the number of species squared.
Then, as an initial study of multicomponent diffusion, we simulate premixed, three-dimensional turbulent hydrogen flames, neglecting secondary Soret and Dufour effects.
Simulation conditions are carefully selected to match previously published results and ensure valid comparison.
Our results show that using the mixture-averaged diffusion assumption leads to a \SI{15}{\percent} under-prediction of the normalized turbulent flame speed for a premixed hydrogen-air flame.
This difference in the turbulent flame speed motivates further study into using the mixture-averaged diffusion assumption for DNS of moderate-to-high Karlovitz number flames.
\end{abstract}

% (Provide 2-4 keywords describing your research. Only abbreviations firmly
% established in the field may be used. These keywords will be used for
% sessioning/indexing purposes.)

\end{frontmatter}

%====================================================================
\section{Introduction}

Implementing full multicomponent mass diffusion transport in direct numerical simulation (DNS) can be memory intensive and computationally expensive.
This is because calculating diffusion fluxes requires point-wise knowledge of the multicomponent diffusion coefficient matrix, which scales with the number of chemical species squared \cite{Bird1960}.
The unity Lewis number, non-unity Lewis number, and mixture-averaged diffusion assumptions have been used to reduce the costs associated with mass diffusion by approximating the full diffusion coefficient matrix as a constant scalar value, a constant vector, and a matrix diagonal, respectively.
In addition, several approaches further reduce the system's complexity by approximating multicomponent diffusion processes in terms of equivalent Fickian processes, such as those used by Warnatz~\cite{Warnatz1978CalculationFlames} and Coltrin et al.~\cite{Coltrin1986ADeposition}.
However, to our knowledge, the accuracy and appropriateness of these assumptions have not been evaluated in turbulent reacting flows against multicomponent diffusion transport due to its high computational expense and a dearth of affordable computing tools.
% To alleviate this cost, and facilitate evaluation of common diffusion assumptions we propose a fast, low cost, stable memory algorithm coupled with a semi-implicit preconditioned iterative method for the time-integration of multicomponent diffusion transport.

As further motivation for this study, Lapointe and Blanquart~\cite{Lapointe2016FuelFlames} recently investigated the impact of differential diffusion on simulations using unity and nonunity Lewis number approximations.
They reported that methane, \textit{n}-heptane, iso-octane, and toluene flames have similar normalized turbulent flame speeds and fuel burning rates when neglecting differential diffusion, but flames using the nonunity Lewis number approximation underpredict the normalized flame speed when including differential diffusion due to reduced burning rates \cite{Lapointe2016FuelFlames}.
Building on these results, Burali et al.~\cite{Burali2016AssessmentFlows} evaluated the relative accuracy of the nonunity Lewis number assumption relative to mixture-averaged diffusion for lean, unstable hydrogen/air flames; lean, turbulent n-heptane/air flames; and ethylene/air coflow diffusion flames.
They demonstrated that the relative error associated with the nonunity Lewis number assumption could be minimized with careful selection of the Lewis number vector for a wide range of flames~\cite{Burali2016AssessmentFlows}.
Similarly, Schlup and Blanquart~\cite{Schlup2018ctm} examined the impact of multicomponent thermal diffusion on DNS of turbulent, premixed, high-Karlovitz hydrogen/air flames.
They showed that simulations using the mixture-averaged thermal diffusion assumption underpredict the normalized flame speeds compared with results from simulations using full multicomponent thermal diffusion.
In addition, including multicomponent thermal diffusion results in increased production of chemical source terms in regions of high positive curvature~\cite{Schlup2018ctm}.
These observed discrepancies in similar flame simulations with different diffusion models warrant a detailed investigation of the fundamental transport phenomena involved.

%%%%% Better for CnF paper
While data are sparse from three-dimensional reacting-flow simulations with multicomponent transport, several groups have investigated the effects of multicomponent transport in simpler configurations.
These studies include one-dimensional~\cite{Coffee:1981,Warnatz:1982, Ern:1999, Bongers:2003, Xin:2015,FAGHIH2018175} and two-dimensional flames~\cite{Charentenay:2002,Giovangigli2015MulticomponentFlames,Dworkin2009} at various unburnt conditions. 
These works compared the multicomponent model with various levels of diffusion and transport property models, from constant Lewis number to mixture-averaged properties.
In general, prior studies found some errors between multicomponent and mixture-averaged formulations for simplified hydrogen/air and methane/air flame configurations, such as unstretched laminar flames.
However, these studies did not assess flames where diffusion effects may be more important, such as two- and three-dimensional, unsteady laminar and turbulent flames.
Moreover, advancing clean and efficient combustion technology requires incorporating realistic fuel chemistry in large-scale turbulent simulations relevant to practical applications. 
Thus, there is a clear need for a computationally efficient algorithm capable of modeling full multicomponent diffusion transport~\cite{LU2009192}.

The studies by Lapointe and Blanquart~\cite{Lapointe2016FuelFlames}, Burali et al.~\cite{Burali2016AssessmentFlows}, and Schlup and Blanquart~\cite{Schlup2018ctm} each took care to isolate the diffusion assumptions in question by neglecting higher-order terms that may affect diffusion transport.
For example, with the exception of Schlup and Blanquart~\cite{Schlup2018ctm}, these studies neglected Soret and Dufour diffusion, as it would be difficult to determine the direct cause of an observed effect when including both molecular and thermal diffusion.
However, despite this methodical approach, the results of these studies were presented with reference to mixture-averaged diffusion, rather than full multicomponent diffusion.
This further highlights the need for a computationally efficient method for implementing full multicomponent transport, and a subsequent examination of the differences between its ``true'' results and those resulting from the approximations conventionally used.

In this direction, several studies have examined the impact of full multicomponent transport on simplified three-dimensional flame configurations.
Giovangigli~\cite{Giovangigli2015MulticomponentFlames} demonstrated that multicomponent Soret effects significantly impact a wide range of laminar hydrogen/air flames.
Specifically, they noted that multicomponent Soret effects influence laminar flame speeds and extinction stretch rates for flat and strained premixed flames, respectively.
For high-pressure systems, Borchesi and Bellan~\cite{Borghesi2015iAConditions} developed and analyzed a multi-species turbulent mixing model for large-eddy simulations.
They focused on turbulent crossflow mixing of a five-species combustion-relevant mixture of \textit{n}-heptane, \ce{O2}, \ce{CO2}, \ce{N2}, and \ce{H2O}.
The multi-species transport model significantly improves the accuracy and fidelity of the solution throughout the mixing layer; however, this study only considered non-reacting flows and, as a result, did not assess the impact of multicomponent transport on the chemistry inherent in turbulent combustion. 
In addition, these simulations implement a simplified diffusion model to approximate multicomponent diffusion but do not directly solve the diffusion terms present in the generalized conservation equations for species and energy \cite{masibellan2013}.

Motivated by the dearth of affordable three-dimensional multicomponent transport models, Ern and Giovangigli~\cite{Ern1995,Ern:1998,Ern:1999} developed the computationally efficient Fortran library EGLIB for accurately determining transport coefficients in gas mixtures.
More recently, Ambikasaran and Narayanaswamy~\cite{Ambikasaran2017AnVelocities} proposed an efficient algorithm to compute multicomponent diffusion velocities, which scales linearly with the number of species.
This significantly reduces computational cost compared with previous methods that directly invert the Stephan--Maxwell equations and thus scale with the number of species cubed.
Although both libraries reduce the computational cost of determining the multicomponent diffusion coefficients, they do not provide a method for reducing the resulting large memory requirements for multidimensional simulations.

Overall, these prior studies provide compelling evidence that multicomponent transport is important and can affect the accuracy of combustion models.
However, none assessed how multicomponent transport impacts three-dimensional turbulent systems with detailed chemistry.
%As noted previously, one- and two-dimensional simulations and non-reacting flows are often cited as a means of computational cost reduction.
In this article, we demonstrate and analyze an efficient, dynamic algorithm that reduces the computational expense of calculating the multicomponent diffusion fluxes.
We demonstrate the approach is accurate and stable for a wide range of time-step sizes.
% We demonstrate the model is second order accurate in time and stable for a wide rang of timesteps.
In addition, we present a comprehensive assessment of the numerical costs associated with this method.
To verify the proposed algorithm we present one-dimensional freely propagating, laminar hydrogen/air flames and compare with results from Cantera.
Finally, we simulate three-dimensional, turbulent, premixed, hydrogen/air flames.
As a preliminary comparison of the mixture-averaged and multicomponent diffusion models, we perform an a posteriori assessment of how the choice of diffusion model impacts the turbulent statistics of the three-dimensional hydrogen simulation.
% perform a priori analysis of the relative direction and magnitude of species flux vectors to assess the relative differences between mixture-averaged and multicomponent mass diffusion transport.

\section{Governing equations}
This section presents the low-Mach number reacting Navier--Stokes equations used in this study.
%governing equations for variable-density,
In addition, this section outlines the method used to determine the mass diffusion fluxes for both the mixture-averaged and multicomponent approaches, abbreviated here as MA and MC, respectively.

\subsection{Low Mach-number equations}

In this work we solve the variable-density, low-Mach number, reacting-flow equations~\cite{Desjardins2008,Savard2015AChemistry}.
The conservation equations are
{\allowdisplaybreaks \begin{IEEEeqnarray}{rCl}
\frac{\partial\rho}{\partial t}+\nabla\cdot\left(\rho \textbf{u}\right) &=& 0 \;, \label{eq:continuity} \\
\frac{\partial\rho \textbf{u}}{\partial t}+\nabla\cdot(\rho\textbf{u}\otimes\textbf{u}) &=& -\nabla p+\nabla\cdot\boldsymbol{\tau} \;, \label{eq:momentum} \\
\frac{\partial\rho T}{\partial t}+\nabla\cdot(\rho \textbf{u}T) &=& \nabla\cdot(\rho\alpha\nabla T)+\rho\dot{\omega}_{T}-\frac{1}{c_{p}}\sum_{i} c_{p,i}\textbf{j}_{i} \cdot \nabla T+\frac{\rho\alpha}{c_{p}}\nabla{c_{p}}\cdot\nabla T \;, \label{eq:energy} \IEEEeqnarraynumspace \\
% 
%double check equation
\frac{\partial\rho Y_{i}}{\partial t}+\nabla\cdot(\rho \textbf{u} Y_{i}) &=& -\nabla\cdot \textbf{j}_{i}+\dot{\omega_{i}} \;, \label{eq:species}
\end{IEEEeqnarray}}%
where $\rho$ is the mixture density, $\textbf{u}$ is the velocity vector, $p$ is the hydrodynamic pressure, $\boldsymbol{\tau}$ is the viscous stress tensor, $T$ is the temperature, $\alpha$ is the mixture thermal diffusivity, $c_{p,i}$ is the constant-pressure specific heat of species $i$, $c_{p}$ is the constant-pressure specific heat of the mixture, $\textbf{j}_{i}$ is the diffusion flux of species $i$, $Y_{i}$ is the mass fraction of species $i$, and $\dot{\omega_{i}}$ is the production rate of species $i$.
In Equation~\eqref{eq:energy}, the temperature source term $\dot{\omega}_{T}$ is given by
\begin{equation} \label{5}
\dot{\omega}_{T}=-c_{p}^{-1}\sum_{i} h_{i}(T)\dot{\omega_{i}} \;,
\end{equation}
where $h_{i}(T)$ is the specific enthalpy of species $i$ as a function of temperature.
The density is determined from the ideal gas equation of state
\begin{equation} \label{6}
\rho=\frac{P_{o}W}{RT} \;,
\end{equation}
where $P_{o}$ is the thermodynamic pressure, $R$ is the universal gas constant, and $W$ is the mixture molecular weight determined via $W=\left(\sum_{i}^N Y_{i}/W_{i}\right)^{-1}$, where $W_{i}$ is the molar mass of the $i$th species and $N$ is the number of species.

The diffusion fluxes are calculated with either the mixture-averaged \cite{Bird1960} or multicomponent \cite{Hirschfelder1954} models, which are both based on Boltzmann's equation for the kinetic theory of gases \cite{Curtiss1949TransportMixtures, Hirschfelder1954}.
The baro-diffusion term is commonly neglected in reacting-flow simulations under the low Mach-number approximation~\cite{Grcar2009TheFlames}.
We have also neglected thermal diffusion because our objective in this work is to investigate the impact of mass diffusion models; Schlup and Blanquart~\cite{Schlup2018ctm,Schlup2018cnf} previously explored the effects of thermal diffusion modeling.
 
\subsection{Mixture-averaged (MA) species diffusion flux}
The $i$th species diffusion flux for the mixture-averaged diffusion model is related to the species gradients by a Fickian formulation and is expressed as
\begin{equation}
    \textbf{j}_{i}= -\rho D_{i,m}\frac{Y_i}{X_i}\nabla X_{i}+\rho Y_{i}\textbf{u}_{c}' \;,
\end{equation}
where $X_i$ is the $i$th species mole fraction, $D_{i,m}$ is the $i$th species mixture-averaged diffusion coefficient as expressed by Bird et al.~\cite{Bird1960}:
\begin{equation}\label{8}
D_{i,m}=\frac{1-Y_{i}}{\sum_{j\neq i} X_{j}/\mathcal{D}_{ji}} \;,
\end{equation}
where $\mathcal{D}_{ji}$ is the binary diffusion coefficient between the $i$th and $j$th species.
Finally, $\textbf{u}_{c}'$ is the correction velocity used to ensure mass continuity:
\begin{equation} \label{MA_correction_velocity}
\textbf{u}_{c}'=\sum_{i=1}^{N} D_{i,m}\frac{Y_{i}}{X_{i}}\nabla X_{i} \;.
\end{equation}
The expression for species diffusion flux can be re-stated in terms of mass fraction $Y_{i}$ as
\begin{equation}\label{7}
\textbf{j}_{i}= - \rho D_{i,m}\left(\nabla{Y_{i}}-Y_i \sum_{k=1}^N \nabla{Y_k}\frac{W}{W_k}\right)+\rho Y_{i}\textbf{u}_{c}' \;,
\end{equation}
where $D_{i,m}$ corresponds to the $i$th element of the diagonal mixture-averaged diffusion coefficient matrix, defined herein as $\dma$. 
%, where $D^{MA}$ is a matrix sized by the number of species squared.

\subsection{Multicomponent (MC) species diffusion flux}
\label{sec:MC_diff}
The multicomponent diffusion model for the $i$th species diffusion flux is
\begin{equation}\label{MC_flux}
    \textbf{j}_{i}=\frac{\rho Y_{i}}{X_{i}W}\sum_{k=1}^N W_{k} D_{ik}\nabla{X_{k}}  \;,
\end{equation}
where $D_{ik}$ is the ordinary multicomponent diffusion coefficient (computed using the \texttt{MCMDIF} subroutine of CHEMKIN II~\cite{Kee1989Chemkin-II:Kinetics} with the method outlined by Dixon--Lewis~\cite{Dixon-Lewis1968FlameSystems}).
Equation~\eqref{MC_flux} can be restated in terms of mass fraction as
\begin{equation}\label{10}
    \textbf{j}_{i}= \rho \sum_{k}-D^{\text{MC}}_{ik}\nabla{Y_{k}} \;, 
\end{equation}
where
\begin{equation}\label{D_MC}
    D^{\text{MC}}_{ik} = -\frac{W_{i}}{W}\left[D_{ik}-\frac{W}{W_{k}}\left(\sum_{j}D_{ij}Y_{j}\right)\right] \;.
\end{equation}
The diagonal of the ordinary multicomponent diffusion matrix, $D_{ii}$, is zero.  
As will be shown later, the $\dmc$ matrix is singular with a kernel of dimension one.  
Interestingly, the vector of species mass fractions is in the kernel:
\begin{equation}\label{dmc_kernel}
    \sum_{k=1}^{N} D^{\text{MC}}_{ik} Y_k = -\frac{W_{i}}{W}\left[\left(\sum_k D_{ik}Y_k\right)-\left(W\sum_k\frac{Y_k}{W_{k}}\right)\left(\sum_{j}D_{ij}Y_{j}\right)\right] = 0\,.
\end{equation}
This property will be important later (in Section~\ref{sec:method_stability}) for the stability analysis.

The multicomponent diffusion coefficients, thermal conductivities, and thermal diffusion coefficients are computed by solving a system of equations defined by the $L$ matrix, composed of nine sub-matrices:
\begin{equation}
\begin{bmatrix}
\textbf{L}^{00,00} & \textbf{L}^{00,10} & 0 \\
\textbf{L}^{10,00} & \textbf{L}^{10,10} & \textbf{L}^{10,01} \\
0 & \textbf{L}^{01,10} & \textbf{L}^{01,01}
\end{bmatrix}
\begin{bmatrix}
\textbf{a}_{1}^{00} \\
\textbf{a}_{1}^{10} \\
\textbf{a}_{1}^{01} \\
\end{bmatrix}
=
\begin{bmatrix}
0 \\
\textbf{X} \\
\textbf{X}
\end{bmatrix} \;,
\end{equation}
where the right-hand side is composed of the one-dimensional mole fraction arrays $\textbf{X}$.
Based on this system of equations, the inverse of the
$\textbf{L}^{00,00}$ block provides the multicomponent diffusion coefficients:
\begin{equation}\label{12}
D_{ij}=X_{i}\frac{16T}{25 P}\frac{W}{W_{j}}(q_{ij}-q_{ii}) \;,
\end{equation}
where
\begin{equation}
\textbf{q} = \left( \textbf{L}^{00,00} \right)^{-1} \;.
\end{equation}
The $\textbf{L}^{00,00}$ sub-matrix block is given by
\begin{equation}
L_{ij}^{00,00} = \frac{16 T}{25 P} \sum_{k=1}^{N} \frac{X_k}{W_i\mathcal{D}_{ik}} \left\lbrace W_j X_j (1 - \delta_{i,k}) - W_i X_i (\delta_{i,j} - \delta_{j,k} ) \right\rbrace \;,
\end{equation}
where $\delta_{i,j}$ is the reduced dipole moment corresponding to the $i$th component of the vector of dipole moments.
%$\mathcal{D}_{ik}$ is the binary diffusion coefficient, $m_{j}$ is the molecular mass of the $j$th species, and $\delta_{ij}$
% The mixture-averaged and multicomponent mass diffusion models will be abbreviated herein as MA and MC, respectively.

%These models are compared using one-dimensional steady; two-dimensional unsteady; and three-dimensional, turbulent, premixed DNS of hydrogen/air flames to determine the accuracy and appropriateness of the mixture-averaged diffusion model as an approximation of full multicomponent diffusion.

\section{Methods}
As discussed previously, multicomponent mass diffusion has not yet been incorporated into three-dimensional turbulent flame simulations due to its high computational expense.
This section presents the discretized equations, numerical algorithm, and preconditioner proposed.
The method is based on the semi-implicit time-marching scheme for species mass-fraction fields proposed by Savard et al.~\cite{Savard2015AChemistry}.

% Do we need to include any discussion of other transport properties at this point?
\subsection{Multicomponent model implementation}\label{model_implementation}
%
% To reduce computational expense while maintaining the fidelity of the ordinary multicomponent diffusion coefficient matrix, our approach calculates the diffusion coefficients once per grid-point and uses a simple dynamic algorithm to reduce memory requirements.
% The approach capitalizes on the knowledge that it is not necessary to store all intermediate flux values through the entire computation; only the fluxes co responding to the current, $(i)$, and future, $(i+1)$, spatial grid-points, in each direction, are stored at any time during the calculation of the diffusion source terms. 
% This is possible because the discretized forms of Equation~\eqref{eq:energy} and \eqref{eq:species} are additive and commutative.
% %Taking advantage of the commutative nature of these equations, the diffusion terms are computed dynamically.

% To ensure a stable solution, the scalar and momentum fields are advanced using a modified form of the semi-implicit preconditioning strategy outlined by Savard et al.~\cite{Savard2015AChemistry}.
% This section proposes a preconditioner corresponding to an approximation of the diagonal of the chemical Jacobian and diffusion coefficient matrices.
% The proposed method uses an iterative approach, that is second-order accurate in time, free of lagging errors to maintain stability and reduce error. 
%This performance makes the semi-implicit strategy ideal for the simulation of turbulent flames with stiff chemistry and multicomponent diffusion transport.

% \subsubsection{Overview of scalar field solver}
%
This work was completed using the structured, multi-physics, and multi-scale finite-difference code NGA~\cite{Desjardins2008,Savard2015AChemistry}.
NGA can solve a wide range of problems, including laminar and turbulent flows~\cite{Xuan2014,Xuan2015,Xuan2016}, constant- and variable-density flows~\cite{Desjardins2008,Carroll2013,Verma2014}, large-eddy simulation~\cite{Xuan2015,Mueller2012}, and DNS~\cite{Bisetti2012,Verma2014,Carroll2013}.
NGA discretely conserves mass, momentum, and kinetic energy with an arbitrarily high-order spatial accuracy \cite{Desjardins2008}.

NGA's variable-density flow solver uses both spatially and temporally staggered variables, storing all scalar quantities ($\rho$, $P$, $T$, $Y_{i}$) at the volume centers and velocity components at their respective volume faces \cite{Desjardins2008,Savard2015}.
The convective term in the species transport equation is discretized using the bounded, quadratic, upwind-biased, interpolative convective scheme (BQUICK) \cite{Herrmann2006}.
The diffusion source term is discretized using a second-order centered scheme and the variables are advanced in time using a second-order semi-implicit Crank--Nicolson scheme \cite{Pierce2001}.

An iterative procedure is applied to fully cover the nonlinearities in the Navier--Stokes equations and the species diffusion terms.
Prior studies demonstrated this iterative process to be critically important for stability and accuracy \cite{Savard2015AChemistry,Desjardins2008,Shunn2012VerificationSolutions,Pierce2001}.
Savard et al.~\cite{Savard2015AChemistry} fully detailed the numerical algorithm sequence; we summarize this method here.
This summary is independent of the preconditioning strategy employed in NGA,
to which propose modifications in Section~\ref{preconditioning}.

The algorithm for advancing one time step follows, using
a uniform time-step size $\Delta{t}$.
The density, pressure, and scalar fields are advanced from time level $t^{n+1/2}$ to $t^{n+3/2}$, and the velocity fields are advanced from time $t^{n}$ to $t^{n+1}$, where $t^{n}$ is the current time.
Each iteration (i.e., time step) consists of $Q$ sub-iterations and follows this procedure:
\begin{enumerate}
\setcounter{enumi}{-1}
    \item Upon convergence of the previous time step, the algorithm stores the density ($\rho^{n+1/2}$), pressure ($P^{n+1/2}$), velocity fields ($\textbf{u}^n$), and scalar fields ($\textbf{Y}^{n+1/2}$),
    where $\textbf{Y}$ represents the vector of species mass fractions $(Y_1, \ldots, Y_N)$.
    The solutions for pressure, species mass fraction, and momentum from the previous time step are used as an initial guess for the iterative procedure: 
    \begin{equation}\label{eq:initial_guesses}
        P_0^{n+3/2} = P^{n+1/2} \;,\;
        \textbf{Y}_0^{n+3/2} = \textbf{Y}^{n+1/2} \;, \text{ and }
        \left(\rho \textbf{u}\right)_0^{n+1} = \left(\rho \textbf{u}\right)^n \;.
    \end{equation}
    An Adams--Bashforth prediction evaluates the initial density:
    \begin{equation}\label{initial_rho}
        \rho^{n+3/2}_0 = 2\rho^{n+1/2}-\rho^{n-1/2} \;,
    \end{equation}
    which ensures that the continuity equation is discretely satisfied at the beginning of the iterative procedure.
    
    \item For the sub-iterations $k=1,\dots,Q$, the semi-implicit Crank--Nicolson method advances the scalar fields in time \cite{Desjardins2008HighFlows,Pierce2001}:
    \begin{equation}\label{15}
        \begin{aligned}
            \rho_{k}^{n+3/2}\textbf{Y}_{k+1}^{n+3/2}=\rho^{n+1/2}\textbf{Y}^{n+1/2}+\Delta{t} \left(\textbf{C}_{k}^{*}+\textbf{Diff}_{k}^{*}+\bm{\Omega}_{k} ^{*}\right) \\
            +\frac{\Delta{t}}{2}\left(\frac{\partial{\textbf{C}}}{\partial{\textbf{Y}}}+\frac{\partial{\textbf{Diff}}}{\partial{\textbf{Y}}}+\frac{\partial{\boldsymbol{\Omega}}}{\partial{\textbf{Y}}}\right)_{k}^{n+1} \cdot \left( \textbf{Y}_{k+1}^{n+3/2}-\textbf{Y}_{k}^{n+3/2} \right) \;,
        \end{aligned}
    \end{equation}
    where $\textbf{Diff}=-\nabla\cdot{\textbf{j}_{i}}$ and $\textbf{Y}_k^*$, $\textbf{C}_k^*$, $\textbf{Diff}_k^*$ , and $\boldsymbol\Omega_k^*$ are the mass fraction, convection, diffusion, and chemical terms evaluated on the mid-point (or half time-step) scalar field $Y_k^*$:
    %
    % \begin{equation}
    \begin{equation}
        \textbf{Y}_{k}^{*} = \frac{\textbf{Y}^{n+1/2}+\textbf{Y}_{k}^{n+3/2}}{2} \;.
        % \boldsymbol{\Omega}_{k}^{*} &= \frac{\boldsymbol{\Omega}^{n+1/2}+\boldsymbol{\Omega}_{k}^{n+3/2}}{2} 
    \end{equation}
    % \end{equation}
    %
    
    %
    % \begin{equation}
    %     \textbf{Diff}=-\nabla\cdot{\textbf{j}_{i}} \;.
    % \end{equation}
    %
    To simplify the discrete notations for spatial differentiation, the operators corresponding to the convective and diffusive terms in Equation~\eqref{eq:species} are written as $\textbf{C}$ and $\textbf{Diff}$, respectively~\cite{Savard2015AChemistry}.
    $\frac{\partial{\textbf{C}}}{\partial{\textbf{Y}}}$ and $\frac{\partial{\textbf{Diff}}}{\partial{\textbf{Y}}}$ are the Jacobian matrices corresponding to the convective and diffusive terms with respect to the species mass fractions, respectively.
    $\textbf{C}$ and $\frac{\partial{\textbf{C}}}{\partial{\textbf{Y}}}$ are functions of the density and velocity, while $\textbf{Diff}$ and $\frac{\partial{\textbf{Diff}}}{\partial{\textbf{Y}}}$ are functions of the density, diffusivity, and molar weight.
    They are consistently updated at each sub-iteration \cite{Savard2015AChemistry}.
   % Convergence of the sub-iterations recovers the fully implicit, second-order time-accurate, Crank--Nicolson formulation.
    \item The density field, $\rho_{k+1}^{n+3/2}$, is evaluated from the new scalar fields using Equation~\eqref{6}. 
    We do not rescale the scalar fields as proposed by Shunn et al.~\cite{Shunn2012VerificationSolutions}.
    However, upon convergence of the sub-iterations, this method is equivalent to the density treatment they proposed~\cite{Savard2015AChemistry}.
    
    \item The momentum equation is advanced in time using a similar semi-implicit Crank--Nicolson method for the scalar fields as described by Savard et al.~\cite{Savard2015AChemistry}.
    
    \item A Poisson equation is then solved for the fluctuating hydrodynamic pressure using a combination of HYPRE~\cite{Falgout2002,Desjardins2008}, BICGSTAB\cite{vanderVorst1992}, and/or FFTW \cite{Frigo2005}.
    The predicted velocity field is then updated.
    
    \item Upon convergence of the sub-iterations, the solutions are updated.
\end{enumerate}
The procedure summarized above becomes equivalent to the fully implicit Crank--Nicolson time-integration scheme upon convergence of the sub-iterations \cite{Pierce2001}.
%Savard et al.~\cite{Savard2015} describe this method in full detail; we direct readers there for additional information on the underlying iterative procedure used in NGA.

\subsection{Preconditioning}
\label{preconditioning}
We expand the above numerical procedure to incorporate multicomponent diffusion by modifying the time-marching step for species mass fraction fields.
%in the method by Savard et al.~\cite{Savard2015AChemistry}.
Specifically, this method modifies the treatment of the mass-diffusion source term in the species mass fraction fields.
All other intermediate steps are unchanged.

\subsubsection{Preconditioning iterative method}
For simpler implementation, Equation~\eqref{15} is solved in its residual form:
\begin{equation}\label{16}
\begin{aligned}
\bigg[\rho_{k}^{n+3/2}\textbf{I}-\frac{\Delta{t}}{2}\bigg(\frac{\partial{\textbf{C}}}{\partial{\textbf{Y}}}+\frac{\partial{\textbf{Diff}}}{\partial{\textbf{Y}}}+\frac{\partial{\boldsymbol{\Omega}}}{\partial{\textbf{Y}}}\bigg)_{k}^{n+1}\bigg] \cdot \left( \textbf{Y}_{k+1}^{n+3/2}-\textbf{Y}_{k}^{n+3/2} \right) \\
= \rho^{n+1/2}\textbf{Y}^{n+1/2}-\rho_{k}^{n+3/2}\textbf{Y}_{k}^{n+3/2}+\Delta{t} \left( \textbf{C}_{k}^{n+1}+\textbf{Diff}_{k}^{n+1}+\bm{\Omega}_{k}^{*} \right) \;.
\end{aligned}
\end{equation}
This equation can be restated as
\begin{equation}
\textbf{Y}_{k+1}^{n+3/2}=\textbf{Y}_{k}^{n+3/2}-\Delta{t}\textbf{J}^{-1} \cdot \Theta_{k} \;,
\end{equation}
where the matrix $\textbf{J}$ is
\begin{equation}
\textbf{J}=\rho_{k}^{n+3/2}\textbf{I}-\frac{\Delta{t}}{2}\bigg(\frac{\partial{\textbf{C}}}{\partial{\textbf{Y}}}+\frac{\partial{\textbf{Diff}}}{\partial{\textbf{Y}}}+\frac{\partial{\boldsymbol{\Omega}}}{\partial{\textbf{Y}}}\bigg)_{k}^{n+1} 
\end{equation}
and the vector
\begin{equation}
\bm{\Theta}_{k}=\frac{\rho_{k}^{n+3/2}\textbf{Y}_{k}^{n+3/2}-\rho^{n+1/2}\textbf{Y}^{n+1/2}}{\Delta{t}}-\left[\textbf{C}_{k}^{n+1}+\textbf{Diff}_{k}^{n+1} +\boldsymbol{\Omega}_{k}^{*} \right]
\end{equation}
is the residual of the species transport equation at the previous sub-iteration, which asymptotes to zero as the sub-iterations fully converge.

Written in its residual form, the time advancement of the species transport equations described here resembles the standard preconditioned Richardson-type iterative method \cite{Savard2015AChemistry,Richardson1911TheDam}, where the matrix $\textbf{J}$ acts as a preconditioner.
The choice of $\textbf{J}$ as a preconditioner is arbitrary and only affects the convergence characteristics of the iterative method \cite{Savard2015AChemistry}.
% does not modify the discrete form of the set of equations to solve.
%Selection of the preconditioner 
For example,
\begin{equation}
\textbf{J}=\rho_{k}^{n+3/2}\textbf{I}
\end{equation}
is equivalent to the fully explicit integration of the convective, diffusive, and chemical source terms in the species transport equations.
Alternatively,
\begin{equation}\label{implicit_jacobian}
\textbf{J}=\rho_{k}^{n+3/2}\textbf{I} - \frac{\Delta{t}}{2}\left(\frac{\partial{\textbf{C}}}{\partial{\textbf{Y}}} + \frac{\partial{\textbf{Diff}}}{\partial{\textbf{Y}}} + \frac{\partial{\boldsymbol{\Omega}}}{\partial{\textbf{Y}}}\right)_{k}^{n+1}
\end{equation}
is equivalent to fully implicit integration of the convective, diffusive, and chemical source terms \cite{Savard2015AChemistry}.

There is a clear tradeoff in selecting the preconditioner.
Since preconditioning is applied to each step of the iterative methods, the form of matrix $\textbf{J}$ should be optimized for low computational and inversion cost while maintaining strong convergence.
The fully explicit preconditioner provides the cheapest option but in our experience results in poor convergence performance, requiring extremely small time steps.
Alternatively, the fully implicit preconditioner would provide excellent convergence criteria and unconditional stability; however, the Jacobian matrices for the chemical and diffusion source terms are typically dense~\cite{Perini2014AMechanisms,Bird1960,Kee1989Chemkin-II:Kinetics}.
Thus, constructing a fully implicit preconditioner is prohibitively expensive for large kinetic models.

To achieve strong convergence while maintaining a low-cost form for the preconditioner, we propose an approximation of the diffusion Jacobian that lies between the fully implicit and fully explicit extremes: a semi-implicit preconditioner.
Savard et al.~\cite{Savard2015AChemistry} previously implemented a similar approach for preconditioning the  chemical Jacobian.

\subsubsection{Semi-implicit preconditioner}
In  Equation~\eqref{implicit_jacobian}, the Jacobian of the diffusion source term depends on the multicomponent diffusion flux, which is proportional to the multicomponent diffusion coefficient matrix, $\dmc$.
However, $\dmc$ is a dense matrix and would be a computationally expensive approximation for the Jacobian.
Alternatively, the mixture-averaged diffusion coefficient matrix, $\dma$, is a simplified approximation of $\dmc$ and thus may provide a reasonable, low-cost approximation of the fully implicit Jacobian.
% Recall that
% \begin{equation}
%     \textbf{Diff}=-\nabla\cdot{\textbf{j}_{i}} \quad \text{and} \quad \textbf{j}_{i}=\rho\sum_{k}D_{ik}^{\text{MC}}-\nabla Y_{k} \;.
% \end{equation}
%
% From mass conservation the sum of the fluxes must be zero;
%
% \begin{equation}
%     \sum_{i} \textbf{j}_{i} = 0 \Rightarrow \sum_{i}\sum_{k}D_{ik}^{MC}\nabla Y_{k} = 0 \Rightarrow \sum_{k}\left(\sum_{i}D_{ik}^{MC}\nabla Y_{k}\right) = 0 \;.
% \end{equation}
%
% knowing that $D_{ik}^{\text{MC}}$ is computed from the local species and temperature values, not global changes, it holds that the diffusion coefficients are independent of global gradients.

The mixture-averaged diffusion coefficient matrix, $\dma$, and the multicomponent diffusion coefficient matrix, $\dmc$, are of a similar order and depend on the underlying species diffusivities.
In addition, since $\dma$ is computed from the local species and temperature values rather than global changes, it is inexpensive to compute.
Finally, since $\dma$ is strictly diagonal and thus inexpensive to invert, it provides a low-cost approximation to the diffusion Jacobian.
In practice the approximate diffusion Jacobian is a tri-diagonal block matrix, where each block is the diagonal $\dma$ matrix.
In other words, for each species the part of the Jacobian corresponding to that species is tri-diagonal and described by $\dma$.

\subsection{Dynamic memory algorithm}\label{memory}
As mentioned previously, high-fidelity simulations with full multicomponent mass diffusion will have a high computational expense.
Thus, to facilitate a cost-effective implementation of full multicomponent diffusion we propose a simple dynamic memory algorithm that significantly reduces the computational resources needed for such simulations.

The cost of simulating full multicomponent diffusion comes from evaluating the $\dmc$ matrix.
%and not the spatial gradients associated with the scalar fields in the conservation equations (Equations~\eqref{eq:energy} and \eqref{eq:species}).
Thus, we can reduce computational cost significantly by limiting the evaluation of $\dmc$ to strictly once per grid-point.
(In contrast, a naive implementation would involve repeated and redundant evaluations when calculating the species diffusion flux vector and its gradient.)
This is possible because the central-difference scheme used is linear and thus additive and commutative by nature.
In other words, the terms in the discretized equation are simply added together, and thus are strictly independent of each other and require no information from the surrounding grid points.

Recognizing this, it follows that the order of addition does not matter so long as all of the appropriate terms are included in the discretization.
Thus, we can calculate the $\dmc$ matrix once per grid point, and calculate and store for each species the discrete terms of the discretized scalar field corresponding only to the information available at that grid point.
The process then repeats at the next grid point and fills in the remaining information.
This approach is simply a memory-efficient rearrangement of the floating-point operations and does not alter the final result.
Moreover, this dynamic memory scheme avoids the need to calculate local gradients at each grid point.

In practice, we calculate and store the portions of the enthalpy and species-diffusion source terms (in Equations~\eqref{eq:energy} and \eqref{eq:species}, respectively) that can be computed from the information available at the $i$th grid-point for the $(i-1/2)$ and $(i+1/2)$ flux vectors.
For example, the discretized form of the diffusion source term is
\begin{IEEEeqnarray}{rCl}
    \textbf{Diff}_i &=& - \nabla \cdot  j_i = \frac{-j_{i+1/2} + j_{i-1/2}}{\Delta{x}} \nonumber \\ 
    &=& \left[\left( \rho_{i} D_{i} + \rho_{i+1} D_{i+1}\right) \frac{Y_{i+1} - Y_{i}}{\Delta{x}} \right. \nonumber \\
    & & -\: \left. \left( \rho_{i-1} D_{i-1} + \rho_{i} D_{i}\right) \frac{Y_{i} - Y_{i-1}}{\Delta{x}}\right] \frac{1}{2\Delta{x}^2} \;, \IEEEeqnarraynumspace
\end{IEEEeqnarray}
where the diffusion source term contributions from the $i-1$, $i$, and $i+1$ grid points are
\begin{align}
    \text{Source}_{i-1} &= \frac{\rho_{i-1} D_{i-1}}{2\Delta{x}^2} \left(Y_{i}-Y_{1-i}\right) \;, \label{i-1_term} \\
    \text{Source}_{i} &= \frac{\rho_{i} D_{i}}{2\Delta{x}^2} \left(Y_{i+1}+Y_{i-1}-2Y_{i}\right) \;, \text{ and} \label{i_term} \\
    \text{Source}_{i+1} &= \frac{\rho_{i+1} D_{i+1}}{2\Delta{x}^2} \left(Y_{i+1}-Y_{i}\right) \;, \label{i+1_term}
\end{align}
respectively.

At the $i$th grid point, information on the diffusion coefficients at the $i-1$ and $i+1$ grid points is not available; thus, only the diffusion coefficients for the $i$th grid point can be stored.
However, by recognizing that $D_i$ at the $i$th grid point is equal to $D_{i+1}$ and $D_{i-1}$ at the $i-1$ and $i+1$ grid points, respectively, it is possible to solve Equation~\eqref{i-1_term}, Equation~\eqref{i_term}, and Equation~\eqref{i+1_term} for the $i+1$, $i$, and $i-1$ grid points, and store them in their respective memory locations. 
At the next grid point ($i+1$) the process repeats and the remaining information for the $i$th grid point is calculated and added to the previously stored partial solution, thus completing the information needed at the $i$th grid point.
Figure~\ref{alg1} summarizes this process; fluxes are located at cell faces while source terms are at cell centers. 
%\textcolor{red}{Note for next paper: This is a pretty critical paragraph, and I think would take a few revisions to make it perfectly clear (maybe with examples).  Not worth worrying about with 3 hours left.}

\begin{figure}[tbh]
\centering
\begin{tikzpicture}
\draw[thick] (0,0) -- (3,0) -- (3,3) -- (0,3) -- (0,0);
\draw[dashed] (0,0) -- (-3,0) -- (-3,3) -- (0,3);
\draw[dashed] (3,0) -- (6,0) -- (6,3) -- (3,3);
\draw[fill=black] (1.5,1.5) circle (0.25em) node[anchor=south] {$i$};
\draw[thick] (-1.5,1.5) circle (0.25em) node[anchor=south] {$i-1$};
\draw[thick] (4.5,1.5) circle (0.25em) node[anchor=south] {$i+1$};
\draw[line width=0.125em,->] (-0.5,0.75) -- (0.5,0.75) node[anchor=north west] {Diff$_{\text{in}}$};
\draw[line width=0.125em,->] (2.5,0.75) -- (3.5,0.75) node[anchor=north west] {Diff$_{\text{out}}$};
\end{tikzpicture}
\\
\begin{algorithm}[H]
\For{i=1:X}{
    Calculate diffusion coefficient matrix\;
    \For{isc=1:N}{
        Flux$(i-1/2) \mathrel{+}= \text{Diff}_{\text{in}, isc}$\;
        Flux$(i+1/2) \mathrel{+}= \text{Diff}_{\text{out},isc}$\;
    }
    $\text{Source}(i) \mathrel{+}=$ influence from $\text{Diff}(i-1/2)$ and $\text{Diff}(i+1/2)$ \;
    $\text{Source}(i-1) \mathrel{+}=$ influence from $\text{Diff}(i-1/2)$\;
    $\text{Source}(i+1) \mathrel{+}=$ influence from $\text{Diff}(i+1/2)$\;
}
\end{algorithm}
\caption{Dynamic algorithm for calculating multicomponent enthalpy and species diffusion source terms. 
Fluxes are located at cell faces while source terms are at cell centers.
$N$ is the number of species.} 
\label{alg1}
\end{figure}

This approach reduces the number of $\dmc$ evaluations from once per species per grid point to strictly once per grid point.
Finally, it reduces temporary memory requirements from an array sized $n_x \times n_y \times n_z \times N^2$ to a $1\times 7$ array corresponding to only the information needed at the current grid point ($i,j,k$) and its six surrounding points,
where $n_x$, $n_y$, and $n_z$ are the numbers of grid points in the $x$, $y$, and $z$ directions.
This optimizes performance by reducing cache calls for both the species mass fractions and species 
diffusion coefficients.

% GB: Removed text because 1) it is not true and 2) it is not necessary
% where the grid spacing needed to calculate the spatial gradients in Equations~\eqref{i-1_term}--\eqref{i+1_term} is simply a scalar quantity, 

The algorithm is most efficient for a structured grid, but the proposed method is easily extendable to finite-volume discretizations on unstructured meshes with scalars located at the cell centers.
In such schemes, the diffusion term is written as the sum of fluxes on each cell surface.
In turn, these fluxes are written as differences of cell-averaged scalar values.
The regrouping of the contributions of the diffusion term to each cell in Equations~\eqref{i-1_term}--\eqref{i+1_term} would follow a similar approach.

\subsection{Method Stability}
\label{sec:method_stability}
To evaluate the theoretical stability of the proposed treatment of the diffusion source terms, we will perform a one-dimensional von Neumann stability analysis.
First, we decompose the vector of species mass fractions into the exact steady-state solution ($\textbf{Y}^{\circ}$) and a small perturbation vector.  Then, we expand this perturbation in a Fourier series by assuming a solution of the form
\begin{equation}
    \textbf{Y}(x,t) = \textbf{Y}^{\circ}(x) + \textbf{f}(t) e^{i\kappa{x}} \;,
    \label{equ:Y_decomp}
\end{equation}
where $\kappa$ is the wavenumber and $\textbf{f}(t)$ is the time-varying amplitude of the perturbation.  Under small deviations from a steady-state solution, we can make the simplifying assumption that
\begin{equation}
    \rho^{n+3/2}_k\approx \rho^{n+1/2}=\rho^{\circ}\;.
\end{equation} 
Similarly, all diffusion coefficients are evaluated from the steady-state solution.

From here, we rewrite Equation~\eqref{15} in a point-wise form neglecting both the chemical source term---demonstrated to be stable by Savard et al.~\cite{Savard2015AChemistry}---and the convective transport term, which is integrated explicitly in this stability analysis (i.e., not modified by sub-iterations).
This transforms the set of $N$ partial differential equations into a set of $N$ ordinary differential equations, where $N$ is the number of species.  Equation~\eqref{16} reduces to the form
\begin{multline}\label{18}
\left(\textbf{I} + \frac{\Delta{t}}{2} \dma\kappa^{\prime 2} \right) \left( \textbf{f}_{k+1}^{n+3/2} - \textbf{f}_{k}^{n+3/2} \right) = \textbf{f}^{n+1/2} - \textbf{f}_{k}^{n+3/2} \\
- \frac{\Delta{t}}{2}\dmc\kappa^{\prime 2} \left( \textbf{f}_{k}^{n+3/2} + \textbf{f}^{n+1/2} \right) \;, 
\end{multline}
where $\kappa^{\prime2}$ is the modified wavenumber, and $\dma$ and $\dmc$ are the mixture-averaged and multicomponent diffusion coefficient matrices calculated from Equations~\eqref{8} and \eqref{D_MC}, respectively.
For the second-order central differencing scheme used, $\kappa^{\prime2}$ takes the form
\begin{equation}\label{modified_wavenumber}
\kappa^{\prime 2} = \frac{2}{\Delta{x}^2} \left[ 1 - \cos(\kappa\Delta{x}) \right] \;.
\end{equation}
While here we apply this to a second-order central difference scheme, the stability analysis holds for any spatial discretization of the diffusion terms in Equation~\eqref{16}.
In the present case, the most unstable mode manifests as cell-to-cell oscillations corresponding to $\kappa=\pi/\Delta x$ and $\kappa^{\prime 2}=4 / \Delta x^2$.

Recall that $\textbf{f}^{n+1/2}$ is the value at the previous time step as defined in step 0 of Section~\ref{model_implementation} and  
\begin{equation}
    \textbf{f}_0^{n+3/2}\equiv \textbf{f}^{n+1/2}\;.
\end{equation}
Dropping the superscripts for clarity, we can reduce Equation~\eqref{18} to
\begin{equation}\label{20}
\textbf{f}_{k+1}=\textbf{A}\textbf{f}_{0}+\textbf{B}\textbf{f}_{k} \;,
\end{equation}
where
\begin{equation} 
\textbf{A} = \left( \textbf{I} + \frac{\Delta{t}}{2}\kappa^{\prime 2}\dma \right)^{-1} \left(\textbf{I} - \frac{\Delta{t}}{2}\kappa^{\prime 2} \dmc \right)
\end{equation}
and
\begin{equation}\label{B_fourier}
\textbf{B} = \left( \textbf{I} + \frac{\Delta{t}}{2}\kappa^{\prime 2}\dma \right)^{-1} \left( \frac{\Delta{t}}{2}\kappa^{\prime 2} \left( \dma - \dmc \right) \right) \;.
\end{equation}
Inspecting Equation~\eqref{20}, matrix $\textbf{A}$ is multiplied by the constant value of the previous time step ($\textbf{f}_{0}$) and therefore does not contribute to the stability of the sub-iterations.
We focus on the properties of the $\textbf{B}$ matrix, which acts as the amplification/growth factor.  Theoretically, the stability of the sub-iterations is ensured if the spectral radius of matrix \textbf{B}, defined as the largest absolute value of the eigenvalues, is less than one:
\begin{equation}
    \rho( \textbf{B} ) \leq 1 \;.
\end{equation}
The matrix $\textbf{B}$ has some interesting properties that deserve further discussion.

\begin{table}[]
    \centering
    \begin{tabular}{@{}cccc@{}}
    \toprule
    $\dmc$\textsuperscript{\textdagger} & $\dma$\textsuperscript{\textdagger} & $\textbf{B}^{\text{imp}}$ & $\textbf{B}^{\text{exp}}$\\\midrule
    0.0000 & 0.2984 & 0.0095 & 0.0000 \\
    0.3033 & 0.3003 & 0.0096 & 25.629 \\
    0.3053 & 0.3014 & 0.0126 & 25.800 \\
    0.3070 & 0.3023 & 0.0131 & 25.945 \\
    0.3860 & 0.4069 & 0.0163 & 32.614 \\
    0.4644 & 0.4585 & 0.0265 & 39.239 \\
    0.4735 & 0.4672 & 0.0276 & 40.012 \\
    1.1163 & 1.0859 & 0.0450 & 94.327 \\
    1.8968 & 1.8145 & \textit{0.9634} & 160.276 \\
    \bottomrule
    \end{tabular}
    \caption{Eigenvalues for the multi-component (left) and mixture-averaged (center-left) diffusion matrices ($\dmc$ and $\dma$) and absolute values of the eigenvalues for the amplification matrix ($\textbf{B}$) for the implicit formulation (center-right) and explicit formulation (right) evaluated on the burned side of the lean hydrogen premixed flame (see section~\ref{sec:1D_flame}).
    \textsuperscript{\textdagger} units are \SI{e-3}{\meter^2\per\second}.
    A time-step size of $\Delta t= $\SI{e-5}{\second} was used for $\textbf{B}$.}
    \label{tab:eig_mc_ma}
\end{table}

First, this matrix is proportional to the difference between the two diffusion matrices $\dma$ and $\dmc$.
Recall that the $\dma$ is a purely diagonal matrix.
Table~\ref{tab:eig_mc_ma} compares the eigenvalues of these matrices on the burned side of the lean hydrogen premixed flames (see Section~\ref{sec:1D_flame} for details on the flame).
The burned side of the flame is characterized by the largest diffusion coefficients and is expected to be the most unstable location within a flame as far as diffusion is concerned.
The two sets of eigenvalues are extremely close, which is expected as the mixture-averaged diffusion model approximates the multi-component diffusion model.
As a result, the norm of the difference of the two matrices is expected to be much less than the norm of either matrix.
In other words, we anticipate that $\rho( \textbf{B} ) \ll 1$ regardless of the time-step size ($\Delta t$) and grid spacing ($\Delta x$).

One noticeable difference between $\dmc$ and $\dma$ is the presence of a null eigenvalue for multi-component diffusion.
As described in Section~\ref{sec:MC_diff}, the multi-component matrix is singular, and its kernel is spanned by the species mass fraction matrix, here $\textbf{Y}^{\circ}$.
Consider the special case of $\textbf{f}_0=\textbf{Y}^{\circ}$.
Leveraging the fact that $\textbf{f}_0$ lies in the kernel of $\dmc$, the amplitude at the next sub-iteration will be
\begin{equation}
    \textbf{f}_{1}=\left(\textbf{A}+\textbf{B}\right)\textbf{f}_{0}= \textbf{f}_{0}\;.
\end{equation}
By recursive reasoning, one can show the property holds for all sub-iterations.
In other words, this mode is unaffected by the iterative process and remains the same between time steps.
This time-invariant mode is nothing more than the steady-state solution $\textbf{Y}^{\circ}$.
To avoid ``double counting'' in Equation~\eqref{equ:Y_decomp}, the eigenvalue analysis of the matrix $\textbf{B}$ should be performed on the linear space not including the vector $\textbf{Y}^{\circ}$.

Table~\ref{tab:eig_mc_ma} provides an example of the eigenvalues of the amplification matrix $\textbf{B}$ for a time-step size of $\Delta t= \SI{e-5}{\second}$.
Practically, the eigenvector associated with the largest eigenvalue of $\textbf{B}$ forms a small angle ($\sim 0.08$ deg) with the species mass fraction vector ($\textbf{Y}^{\circ}$). Hence, most of it is in the kernel of $\dmc$.
Following the previous discussion, this eigenvalue (indicated in italics) should not be considered, and the overall stability is controlled by the second-largest eigenvalue, in this case $0.0425$.
As expected, this eigenvalue is much less than unity, thus proving the stability of the iterative procedure.
This should be compared to the stability of the explicit formulation obtained by setting $\dma=0$ in Eq.~\eqref{B_fourier}.
Under these conditions, the spectral radius of $\textbf{B}$ becomes
\begin{equation}
    \rho(\textbf{B}) = 2\frac{\Delta t}{\Delta x^2}\rho\left(\dmc\right)\approx 2\frac{\Delta t}{\Delta x^2}\max\left(\dma\right)\,,
\end{equation}
which resembles a Fourier number.
As shown by the large eigenvalue of the explicit-method amplification matrix, $\mathbf{B}^{\text{exp}}$, in Table~\ref{tab:eig_mc_ma}, solving the system of equations would not be stable at $\Delta t= \SI{e-5}{\second}$ without the proposed implicit formulation.
Section~\ref{practical_stability} presents an in-depth comparison of this theoretical stability criterion against practical numerical convergence results for a one-dimensional freely propagating flame.

\section{Test cases}

We will evaluate the performance of the proposed iterative method and the relative cost of the implemented memory algorithm in Section~\ref{results}.
We base our evaluation on two flow configurations: a one-dimensional, unstretched, laminar flame and a three-dimensional, statistically stationary, turbulent flame; both are premixed hydrogen/air flames.
All simulations used the same nine-species hydrogen mechanism of Hong et al.~\cite{Hong2011AnMeasurements} with updated rate constants from the same group~\cite{Lam2013AAbsorption,Hong2013OnAbsorption}.
This section describes the configuration and conditions used for the one- and three-dimensional simulations used for this study.
\ref{method_validation} includes additional method verification.

\subsection {One-dimensional premixed flame}
\label{sec:1D_flame}

To verify the implementation of the multicomponent mass-diffusion model and evaluate its accuracy, we performed one-dimensional, unstretched (flat), laminar flame simulations and compared these with similar mixture-averaged and multicomponent results computed using Cantera~\cite{Goodwin2017}.
We selected the one-dimensional flat flame configuration because it restricts all transport to the streamwise direction.
% this simplifies computation and ensures similar magnitudes in the species source terms for all methods.
As a result, the spanwise fluxes are zero by definition for this geometry.
This condition may not hold in a multidimensional flow simulation where the multicomponent diffusion fluxes may be misaligned with the species gradient vector.
This simplified geometry allows us to directly compare the multicomponent mass diffusion model to the commonly used mixture-averaged diffusion model.
% For this comparison, we simulated a one-dimensional unstretched (flat) laminar hydrogen/air flame with an equivalence ratio of $\phi=0.4$ for all Cantera and NGA cases.

The simulations used an unburnt temperature of \SI{298}{\kelvin} and pressure of \SI{1}{atm}, with an equivalence ratio of $\phi=0.4$ and inlet velocity equal to the laminar flame speed for all Cantera and NGA cases.
The flame was centered in a computational domain comprised of 720 grid points where $\Delta x=$\SI{15.4}{\micro\metre}.
To ensure fidelity in the results, we selected the domain to have at least 20 points through the laminar flame, with the thickness defined using
the maximum temperature gradient: $l_{F} = (T_{\max} - T_{\min})/\lvert\nabla{T\rvert_{\max}}$.
Schlup and Blanquart~\cite{Schlup2018ctm} used an identical configuration to investigate the impact of Soret and Dufour thermal diffusion effects.

We ran the Cantera simulations similarly using both mixture-averaged and multicomponent diffusion models with matching inlet conditions, equivalence ratio, and domain size.
The freely-propagating adiabatic flat flame solver (\texttt{FreeFlame}) was used with grid refinement criteria for both slope and curvature set to 0.1 and a refinement ratio of 2.0 for 860 grid-points.
%Finally, the Cantera simulations were solved to steady state with the energy equation enabled.

\subsection{Three-dimensional flow configuration}
We simulated a three-dimensional, turbulent, premixed, freely propagating flame
as a test of the proposed algorithm for multicomponent mass diffusion and to assess the impact of diffusion model choice on global statistics such as the turbulent flame speed.
The computational domain consists of inflow and convective outflow boundary conditions in the streamwise direction.  
The two spanwise directions use periodic boundaries. 
We set the inflow velocity to the mean turbulent flame speed, which keeps the flame statistically stationary such that turbulent statistics can be collected over an arbitrarily long run time.
%Table~\ref{tab:3D_flow_config} also gives details of these computational domains.
In the absence of mean shear, we use a linear turbulence-forcing method \cite{Rosales2005,Carroll2013} to maintain the production of turbulent kinetic energy through the flame.
We carefully selected the Karlovitz number to fall within the distributed reaction zone regime while avoiding the broken reaction zone regime~\cite{Lapointe:2015}.
Moreover, the computational setup for this case is similar to those of Lapointe et al.~\cite{Lapointe:2015}, Burali et al.~\cite{Burali2016AssessmentFlows}, and Schlup et at.~\cite{Schlup2018ctm}, who studied  differential-diffusion effects, local extinction, and flame broadening using the mixture-averaged model and constant non-unity Lewis number assumptions.

\begin{table}[htbp]
\small
    \caption{Three-dimensional simulations parameters. $\Delta x$ is the grid spacing, $\eta_{u}$ is Kolmogorov length scale of the unburnt gas, $\Delta t$ is the simulation time-step size, $\phi$ is the equivalence ratio, 
    $P_0$ is the thermodynamic pressure, $T_u$ is the temperature of the unburnt mixture, 
    $T_{\text{peak}}$ is the temperature of peak fuel consumption rate in the one-dimensional laminar flame, $S_L$ is the laminar flame speed, $l_F = \left(T_b - T_u\right)/\left|\nabla T\right|_{\max}$ is the laminar flame thickness, $l = u'^3/\epsilon$ is the integral length scale, $u'$ is the turbulence fluctuations, $\epsilon$ is the turbulent energy dissipation rate, $\text{Ka}_u$ is the Karlovitz number of the unburnt mixture, $\text{Re}_t$ is the turbulent Reynolds number of the unburnt mixture, and $\nu_u$ is the unburnt kinematic viscosity.}
    \centering
    %\begin{tabular}{@{\extracolsep{\fill}}l c c c c c c c c@{}}
    \begin{tabular}{@{}l c c@{}}
         \toprule
         % & \multicolumn{2}{c}{3D \ce{H2}} \\
         %\midrule
         Parameter & MA & MC \\
         \midrule
         Domain & \multicolumn{2}{c}{$8L \times L \times L$} \\
         $L$ & \multicolumn{2}{c}{$190\Delta{x}$}  \\
         Grid & \multicolumn{2}{c}{$1520 \times 190 \times 190$} \\
         $\Delta{x}$ [\si{\m}] & \multicolumn{2}{c}{\num{4.24e-5}} \\
         $\eta_{u}$ [\si{\m}] & \multicolumn{2}{c}{\num{2.1e-5}} \\
         $\Delta{t}$ [\si{\s}] & \multicolumn{2}{c}{\num{6e-7}} \\
         $\phi$ & \multicolumn{2}{c}{0.4} \\
         $P_0$ [\si{\atm}] & \multicolumn{2}{c}{\num{1}} \\
         $T_{\text{u}}$ [\si{\K}] & \multicolumn{2}{c}{\num{298}} \\
         $T_{\text{peak}}$ [\si{\K}] & 1190 & 1180 \\
         $S_{L}$ [\si{\m\per\s}] & 0.230 & 0.223 \\
         $l_{F}$ [\si{\mm}] & 0.643 & 0.631 \\
         $l/l_{F}$ & 2 & 2.04 \\
         $u'/S_{L}$ & 18 & 18.6 \\
         $\text{Ka}_u = \tau_{F}/\tau_{\eta}$ & 149 & 151 \\
         $\text{Re}_t = (u'l)/\nu_{u}$ & \multicolumn{2}{c}{289} \\
         \bottomrule
    \end{tabular}
    \label{tab:3D_flow_config}
\end{table}

Table~\ref{tab:3D_flow_config} provides further details of the computational domain, unburnt mixture, and inlet turbulence.
The unburnt temperature and pressure are \SI{298}{\kelvin} and \SI{1}{\atm}, respectively.
The inlet equivalence ratio is $\phi = 0.4$, with an unburnt Karlovitz number $\text{Ka}_{u}$ = $\tau_F/\tau_{\eta} = 149$, where $\tau_F = l_F / S_L$ is the flame time scale and
$\tau_{\eta}=(\nu_u /\epsilon)^{1/2}$ is the Kolmogorov time scale of the incoming turbulence with unburnt kinematic viscosity $\nu_u$ and turbulent energy dissipation $\epsilon$. 
The unburnt turbulent Reynolds number is $\text{Re}_{t} = u'l/\nu_u=289$, where $u'$ is the fluctuation of the mean velocity and $l$ is the integral length scale.
The mean inflow velocity at the inlet boundary condition approximately matches the turbulent flame speed so that the flame remains relatively centered in the domain and we can perform arbitrarily long simulations. 
Once the turbulence has fully developed, we run the simulations for 22 eddy turnover times, $\tau_{\text{eddy}} = k/\epsilon\approx \SI{500}{\micro\second}$.

\begin{figure}[htb]
    \centering
        \includegraphics[width=\textwidth]{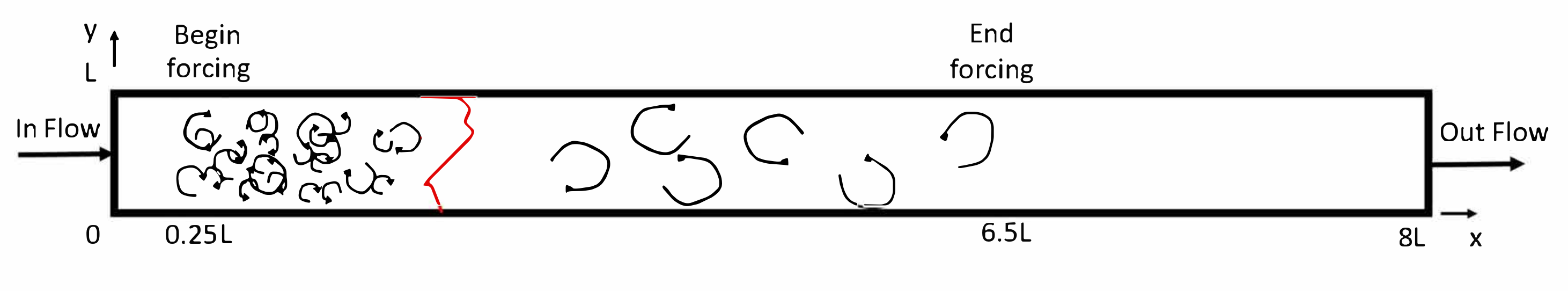}
        \caption{Two-dimensional schematic of the three-dimensional flame configuration. 
        Adapted from Burali et al.\ and Schlup and Blanquart~\cite{Burali2016AssessmentFlows,Schlup2018ctm}.
        The red line indicates the approximate location of the flame.
        } 
        \label{3D_Schematic}
\end{figure}

\begin{figure}[htb] 
  \centering
    \includegraphics[width=0.6\textwidth]{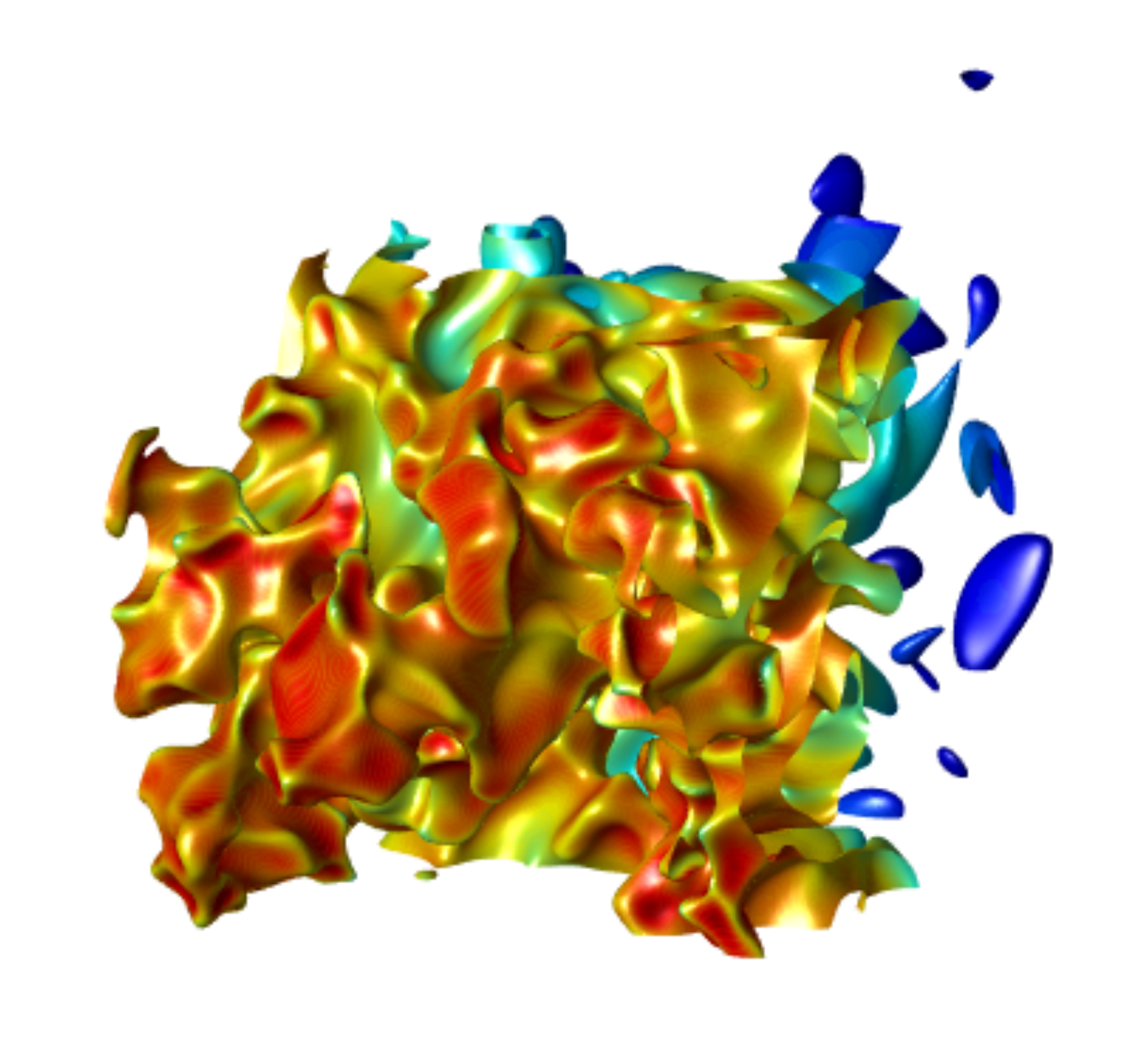}
    \caption{Iso-surface of peak temperature colored by \ce{OH} mass fraction for a three-dimensional turbulent hydrogen/air flame with multicomponent mass diffusion.}
  \label{3D_H2}
\end{figure}

The domain has 1520 points in the streamwise direction and 190 points in both spanwise directions, with a uniform grid size of $\Delta{x}=l_{F}/16$.
This domain is about $100 l_{F}$ long and $12 l_{F}$ in the spanwise directions.
% thus, the spanwise dimensions of this case are approximately 40\% of the two-dimensional freely propagating dimensions.
Given the prescribed turbulence intensity, this mesh has a grid spacing equivalent to $\Delta{x}\approx2\eta_u$, where $\eta_u$ is the Kolmogorov length scale for the unburnt region; this resolution improves in the burnt region of the flame.
Lapointe et al.~\cite{Lapointe:2015} previously confirmed the suitability of the selected grid spacing and resolution in the flame front using a mesh refinement study, which found no difference when using this grid spacing compared with half the size.
Figure~\ref{3D_Schematic} shows a two-dimensional schematic of the domain, including the locations of the flame and the forcing region.
Figure~\ref{3D_H2} shows a three-dimensional view of the iso-surface of $T_{\text{peak}}$ defining the flame front, where $T_{\text{peak}}$ is the temperature of peak fuel consumption rate in the one-dimensional laminar flame. 
The flame surface shows the complex behavior of the flame in the turbulent field.

\section{Results and discussion}\label{results}
% \section{Results and discussion}
%
To start, 
we present a practical assessment of the method's convergence and stability, by comparing the numerical rate of convergence to the theoretical rate of convergence.
Following this demonstration of the proposed method's stability, we verify the accuracy of the method through a posteriori assessment of one-dimensional, unstretched, premixed, laminar flame simulations.
Finally, we present a preliminary evaluation of the relative differences between the mixture-averaged and multicomponent diffusion models for the three-dimensional turbulent premixed flame simulations.

\subsection{Stability analysis results}\label{practical_stability}

We use the one-dimensional flame to numerically evaluate the convergence stability of the sub-iterations with respect to time-step size.
The simulations for these tests were initialized from a mixture-averaged data file to provide a worst-case scenario for the initial iterative step in converging to the multicomponent solution.
While the theoretical analysis was performed assuming explicit transport of the convective terms and constant density\slash diffusion coefficients, we performed this test with semi-implicit transport and variable density\slash diffusion coefficients.
This demonstrates the stability of the proposed preconditioner for the semi-implicit multicomponent diffusion transport in a practical numerical simulation.
Savard et al.~\cite{Savard2015AChemistry} previously showed the numerical stability of the chemical and convective terms, so we do not discuss these terms in detail in this analysis.

\begin{figure}[htb!]
    \centering
    \begin{subfigure}{0.49\textwidth}
        \centering
        %\includegraphics[width=\textwidth]{Convergence_Residual_Stable.pdf}
        %\caption{Stable convergence}
        \includegraphics[width=\textwidth]{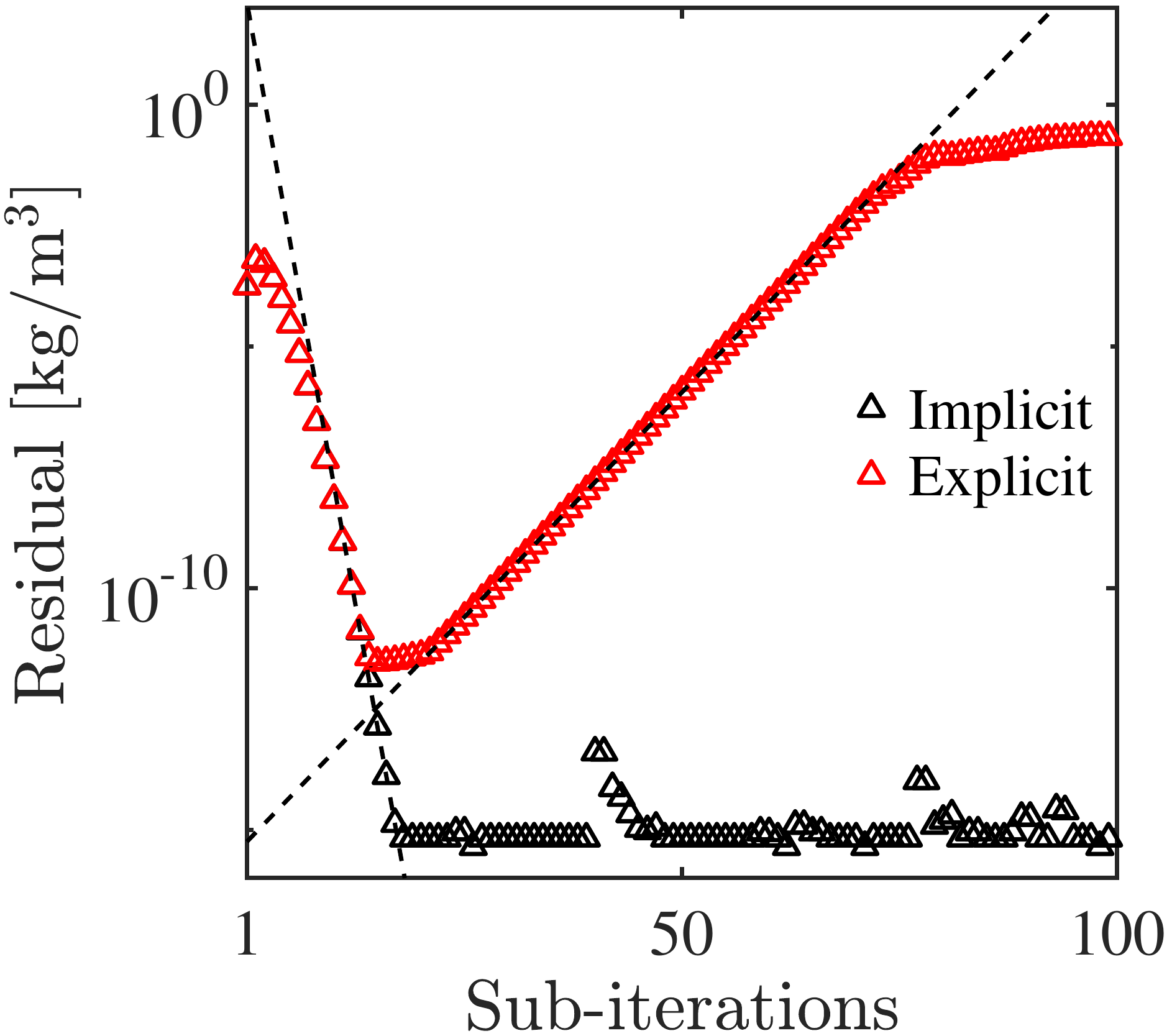}
        \caption{$\Delta t=10^{-7}s$}
        \label{Stability (a)}
    \end{subfigure}
    \hfill
    \begin{subfigure}{0.49\textwidth}
        \centering
        %\includegraphics[width=\textwidth]{Convergence_Residual_Unstable.pdf}
        %\caption{Unstable convergence}
        \includegraphics[width=\textwidth]{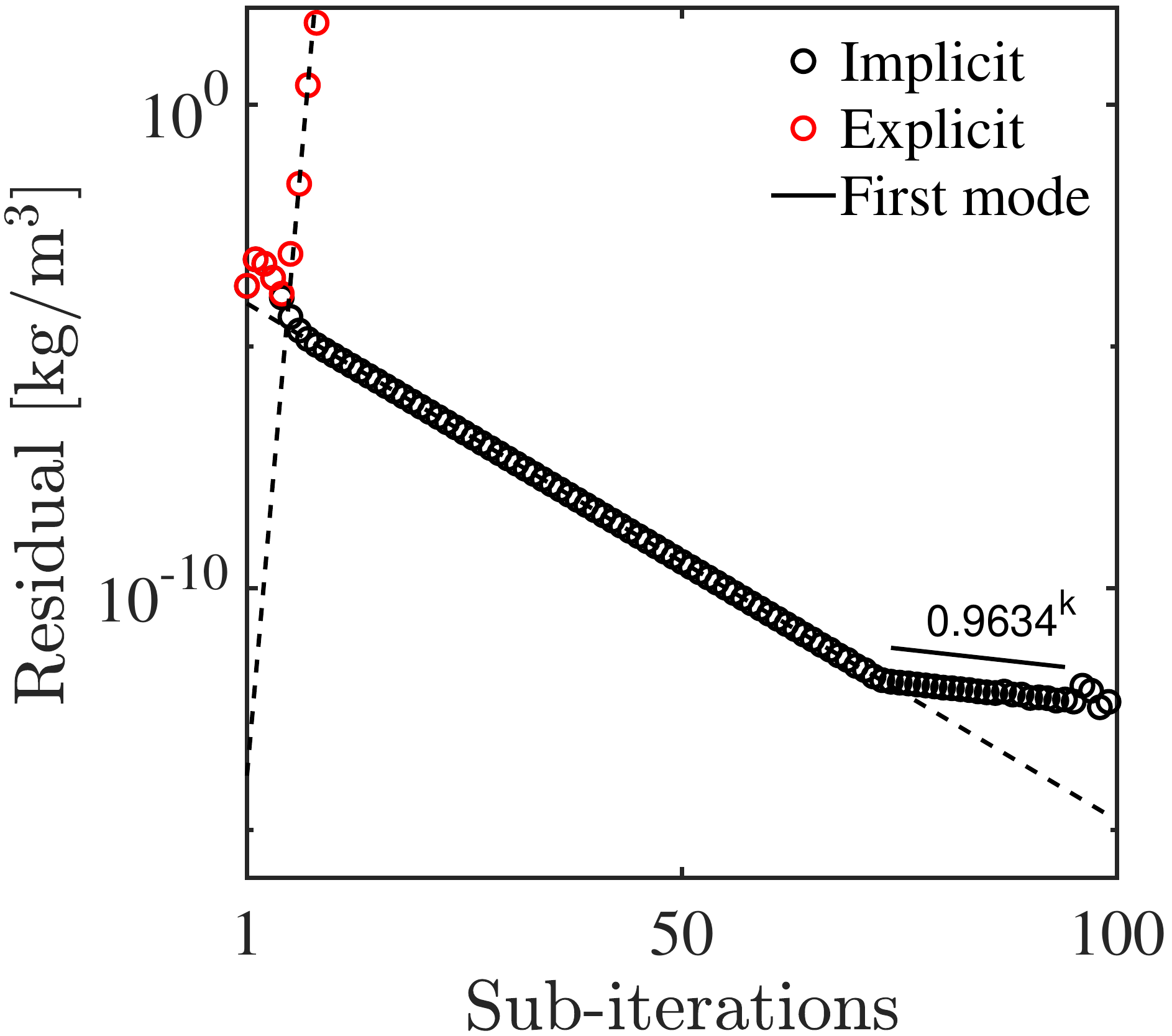}
        \caption{$\Delta t=10^{-5}s$}
        \label{Stability (b)}
    \end{subfigure}
    \caption{Convergence of the density residual as a function of sub-iteration for the proposed semi-implicit method, for a smaller and larger time-step size.
    Dashed lines are the spectral radii shown in Figure~\ref{Convergence_rate} and are determined by numerically fitting an exponential curve to the slope of the density residual.}
    \label{Stability}
\end{figure}

We focus on the maximum density residual over the whole domain, because  its convergence is controlled by the convergence of all chemical species.
Figures~\ref{Stability (a)} and~\ref{Stability (b)} present the density residuals as a function of sub-iteration, starting from the initial time step, for a small and large time-step size, respectively.
For the time-step sizes tested, converging (as opposed to converged) sub-iterations implies a stable simulation, which agrees with behavior shown by Savard et al.~\cite{Savard2015AChemistry}.
In other words, unless the sub-iterations diverge, the simulation remains stable.
As expected, the explicit method diverges quickly even at very small time-step sizes (Figure~\ref{Stability (a)}), while the semi-implicit method remains stable 
up to a time-step size of $\Delta t \leq$ \SI{1e-5}{\second} (Figure~\ref{Stability (b)}).

The rate of convergence of the sub-iterations for each of the source terms in Figures~\ref{Stability (a)} and~\ref{Stability (b)} follows an exponential relationship, i.e., $\text{Res}_k \sim r^k$, where $\text{Res}_k$ is the residual of the $k$th sub-iteration and $r$ is the convergence rate.
We compute the numerical convergence rate $r$ by fitting an exponential curve to the slope of the density residuals; the convergence rate is represented by dashed lines in Figure~\ref{Stability}.
Since density is a function of the species mass fractions, its convergence rate should tend towards that of the slowest-converging species mass fraction.

\begin{figure}[htb!]
    \centering
        \includegraphics[width=0.6\textwidth]{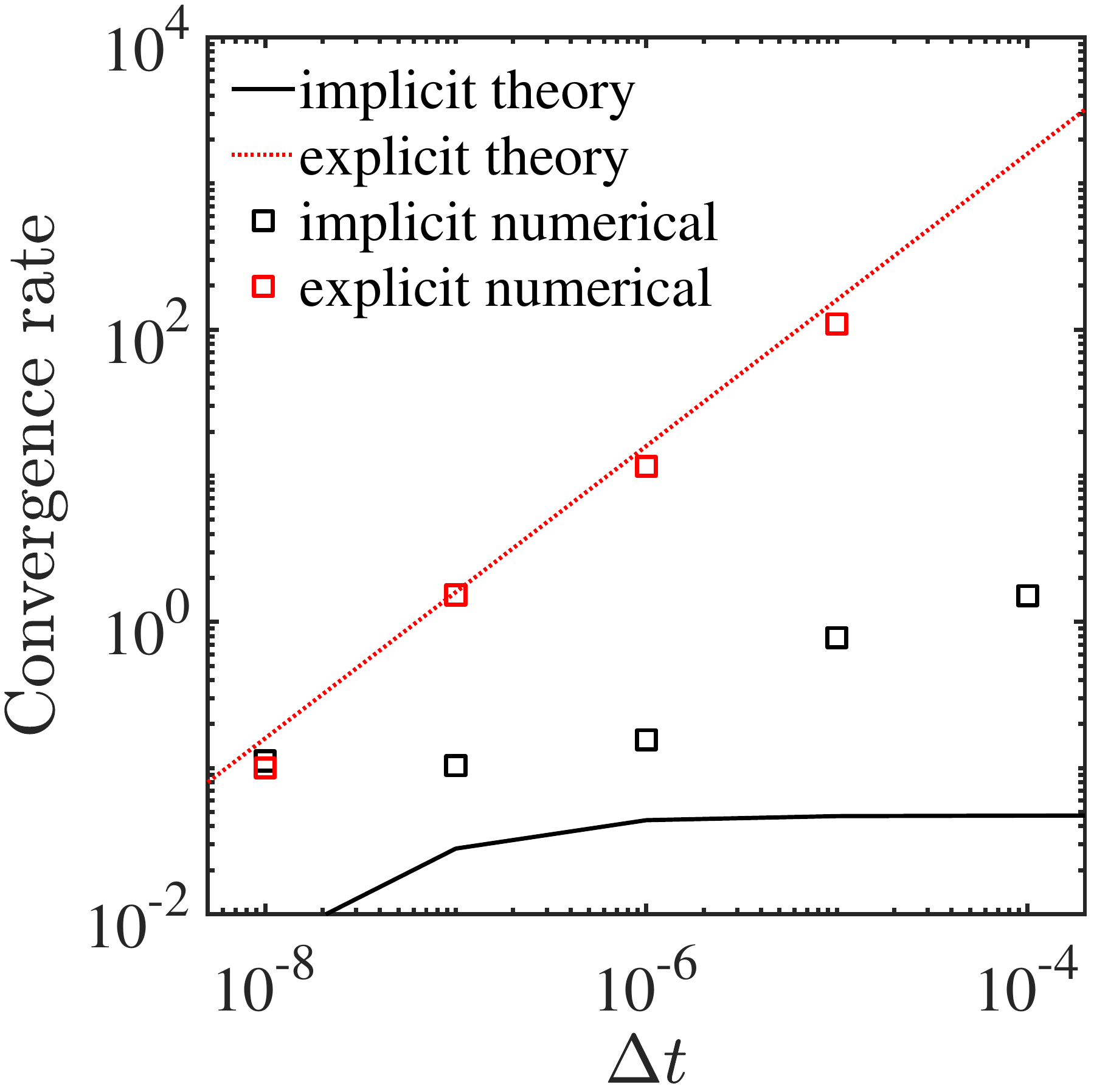}
        \caption{Theoretical convergence rate determined from diagonalizing matrix $\textbf{B}$ corresponding to the worst-case modified wavenumber for the one-dimensional premixed flame, compared with the numerical convergence rates determined by fitting an exponential curve to the slope of the density residual.}
        \label{Convergence_rate}
\end{figure}

Figure~\ref{Convergence_rate} compares the results of the theoretical and numerical stability analyses, showing the spectral radius of matrix $\textbf{B}$ as a function of the time-step size for the one-dimensional test case.
For the explicit scheme, the theoretical and numerical results agree well for the full range of time-step sizes. 
However, for the implicit scheme, the predicted spectral radius is much smaller than the measured one.
The proposed implicit formulation thus yields a convergence rate that is not limited by diffusion, but rather constrained by other processes that were not considered in the stability analysis, such as chemistry.
The predicted convergence rate can nonetheless be observed in the implicit case.
As mentioned in Section~\ref{sec:method_stability}, the eigenvector associated with the largest eigenvalue of $\textbf{B}^{\text{imp}}$ in Table~\ref{tab:eig_mc_ma} forms a small angle with $\textbf{Y}^{\circ}$.
Hence, a fraction of the error, albeit tiny, is associated with this eigenvector, which will slowly converge. 
In Figure~\ref{Stability (b)}, the convergence rate for the last sub-iterations of the implicit case closely matches that eigenvalue.

Overall, these results suggest that the theory well-approximates actual stability and provides a practical limit for the numerical stability of the proposed algorithm.

\subsection{Method verification}

\begin{figure}[htb!] 
  \centering
  \includegraphics[width=1\textwidth]{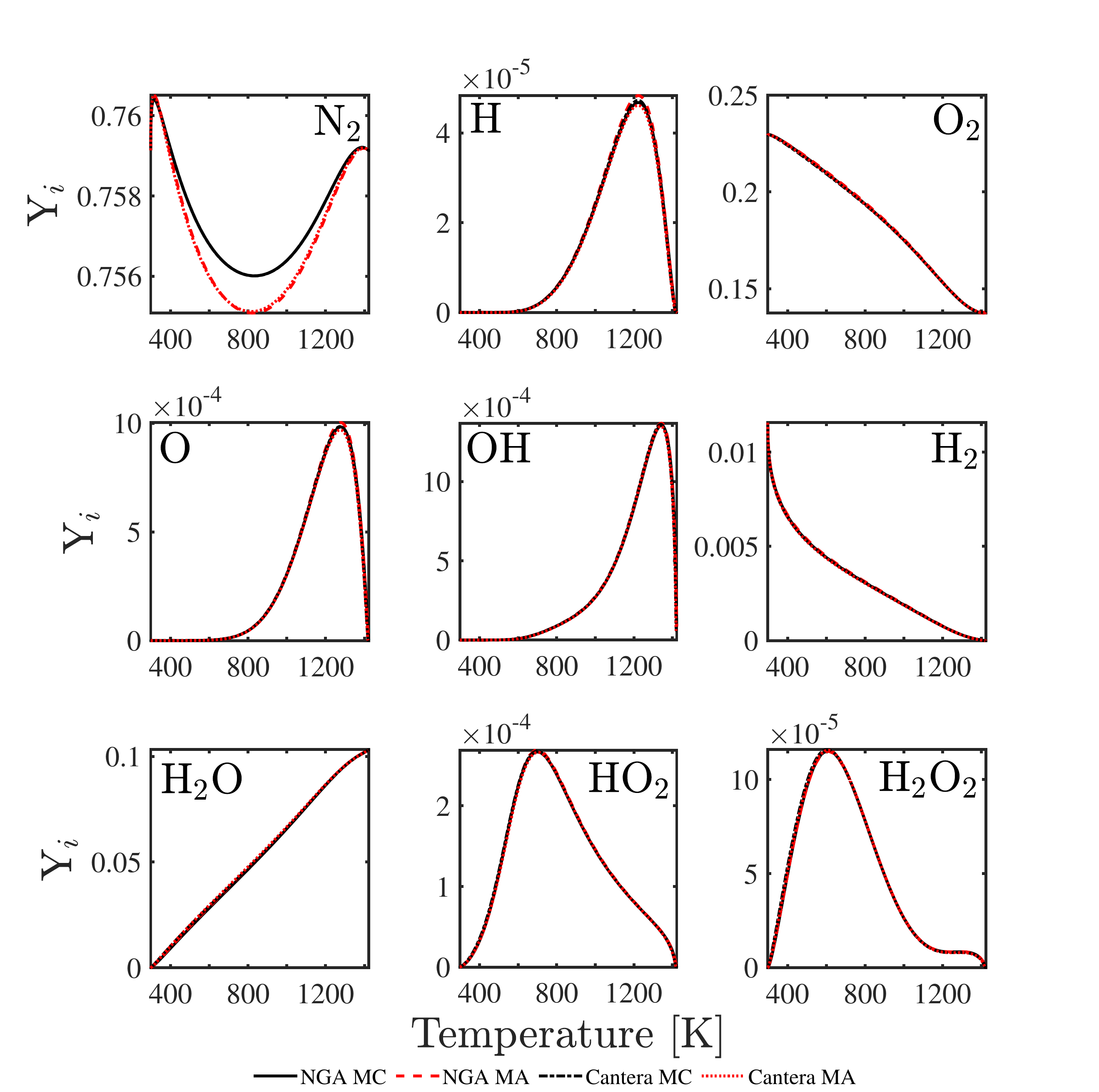}
  \caption{A posteriori comparisons of species mass fractions relative to mixture local temperature in a hydrogen/air flame with $\phi=0.4$ using NGA and Cantera.}
  \label{1D_species}
\end{figure}

To verify the multicomponent model, we compare a posteriori the one-dimensional unstretched species profiles and laminar flame speeds.
Figure~\ref{1D_species} compares the nine species profiles for the steady-state one-dimensional flat flame solutions relative to local mixture temperature for the multicomponent and mixture-averaged models from both NGA and Cantera.
The profiles all agree within \SI{1}{\percent} at all points, with the exception of \ce{N2}.
The laminar flame speeds ($S_{L}^o$) for these simulations are approximately \SI{23.0}{\centi\meter\per\second} and \SI{22.3}{\centi\meter\per\second} for the mixture-averaged and multicomponent diffusion NGA cases, respectively; the laminar flame speeds for both cases agree with those from Cantera within \SI{1}{\percent}. 
The unstretched laminar flame speed is
\begin{equation}
S_{L}^o=-\frac{\int \rho\dot{\omega}_{\ce{H2}}dx}{\rho_{u}Y_{\ce{H2},u}} \;,
\end{equation}
where $\rho_{u}$ is the unburnt mixture density and $Y_{\ce{H2},u}$ is the unburnt fuel mass fraction.
We attribute the larger difference in the species profile for \ce{N2} to the correction velocity term associated with the mixture-averaged diffusion model, which is weighted by mass fraction and thus can be heavily impacted by differences in \ce{N2} due to its high concentration throughout the flame.
The minor differences between the multicomponent species profiles are less than \SI{1}{\percent} at all points.
The strong agreement between the other eight species profiles for both the NGA and Cantera results verifies the multicomponent model's functionality.
%Moreover, the similarity in results between the mixture-averaged and multicomponent cases suggests that the mixture-averaged diffusion assumption is sufficient for one-dimensional flames.

\subsection{Accuracy}

With the proposed algorithm's stability limits and functionality verified, we now examine the accuracy for a given stable simulation.
We determine the order of accuracy of the method based on the 1D freely propagating flame case by examining the power-law dependence of the error as a function of the time-step size.

\begin{figure}[htbp]
  \centering
    \begin{subfigure}{0.49\textwidth}
        \centering
         \includegraphics[width=\textwidth]{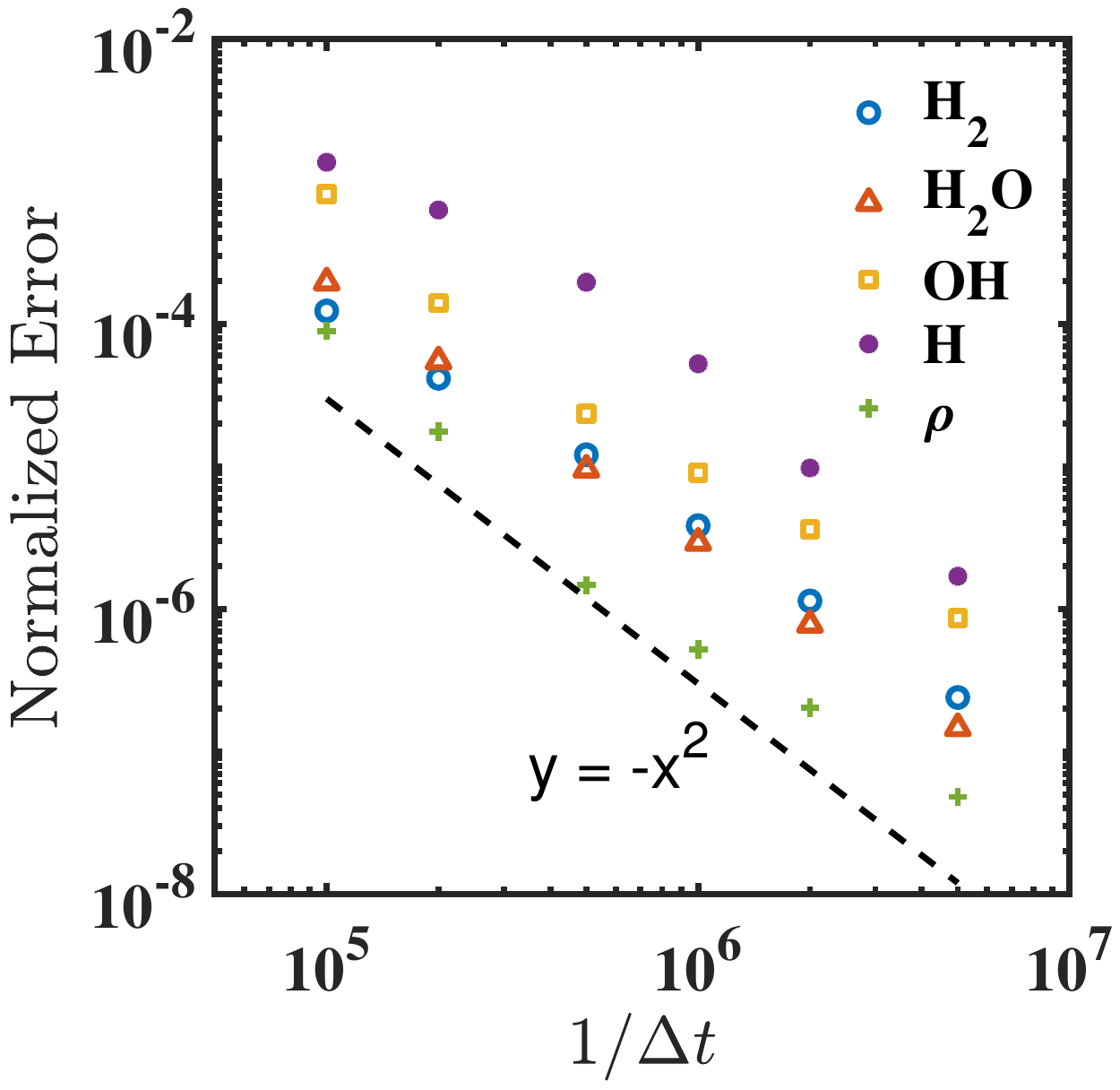}
        \caption{Seven sub-iterations}
        \label{Normalized_error (b)}
    \end{subfigure}\
    \begin{subfigure}{0.49\textwidth}
        \centering
        \includegraphics[width=0.9\textwidth]{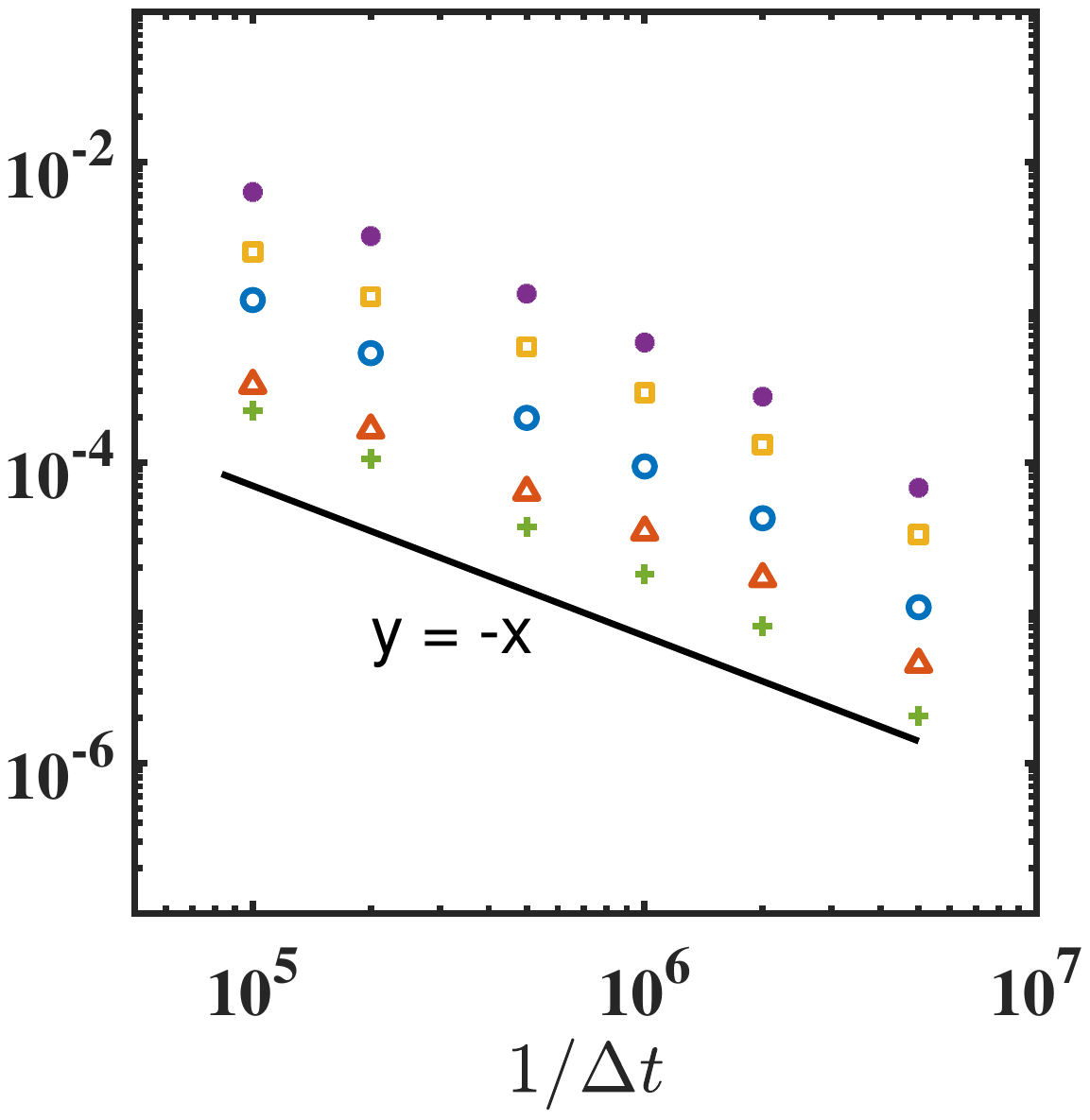}
        \caption{Four sub-iterations}
        \label{Normalized_error (a)}
    \end{subfigure}
  \caption{Relative accuracy of the method as a function of time step size for the one-dimensional, freely propagating flame test case with seven sub-iterations. Errors are defined as the absolute difference of their integrated value in temperature space compared with a reference solution obtained for $\Delta t = \SI{1e-7}{\second}$ and seven sub-iterations. Black dashed line corresponds to $y = x^{-2}$.}
  \label{Normalized_error}
\end{figure}

Figure~\ref{Normalized_error} shows the normalized error for the 1D freely propagating flame case for various time steps.
We initialize the simulation using an input flame profile corresponding to a fully converged statistically stationary flame, generated with a time-step size of $\Delta t = \SI{1e-7}{\second}$ and seven sub-iterations.
A wall is then set at the simulation inlet, allowing the flame to propagate upstream in the domain.
We then let the reference flame propagate for two flame pass-through times to ensure a fully converged freely propagating flame profile free of any initial transients due to the transition from the input stationary flame profile.
This reference file then serves as the input for a set of freely propagating flames with time-step sizes ranging \SIrange{e-5}{e-7}{\second} and for seven sub-iterations.
Finally, we allow these test flames to propagate for an additional flame pass-through time to ensure statistical independence from the initial reference-flame input file.

With the freely propagating flame tests completed, we interpolated the species and density fields to a constant temperature space corresponding to the temperature distribution in the flame region.
This interpolation ensures a direct comparison of the species error independent of variation in the temperature space over the range of time-step sizes.
We then calculate error as the $L^2$-norm of the species and density profiles in temperature space, relative to the reference flame profile with $\Delta t = \SI{1e-7}{\second}$ and seven sub-iterations:
\begin{equation}
    \text{error} = \sqrt{\frac{\int \left(Y_{i}-Y_{i, \text{ref}}\right)^2 dT}{\int Y_{i,\text{ref}}^2 dT}}
\end{equation}
and
\begin{equation}
    \text{error} = \sqrt{\frac{\int \left(\rho-\rho_{\text{ref}}\right)^2 dT}{\int \rho_{\text{ref}}^2 dT}} \;.
\end{equation}
We selected the species \ce{H2}, \ce{H2O}, \ce{OH}, and \ce{H} to evaluate the accuracy of the method because they represent the reactants, intermediate species, and products present in hydrogen combustion.
Density ($\rho$) is also included to globally assess error, since it depends on all species.
As shown in Figure~\ref{Normalized_error (b)}, all quantities exhibit second-order accuracy in time with seven sub-iterations.
The errors corresponding to the $L^1$- and $L^{\infty}$-norms are similar in magnitude and also demonstrate second-order accuracy in time with seven sub-iterations.

While the method is fully second-order accurate for seven sub-iterations and above, the solution transitions to first-order accuracy as the number of sub-iterations decreases.
Figure~\ref{Normalized_error (a)} shows that the solution exhibits first-order accuracy when using four sub-iterations.
Between four and seven sub-iterations the solution is second-order accurate for large time-step sizes but transitions to first-order accuracy as the time step size decreases.
The range of time-step sizes that achieve second-order accuracy grows until the solution becomes fully second-order accurate
at seven sub-iterations for all time-step sizes considered.

\begin{figure}[htb!]
    \centering
    \begin{subfigure}{0.49\textwidth}
        \centering
        \includegraphics[width=\textwidth]{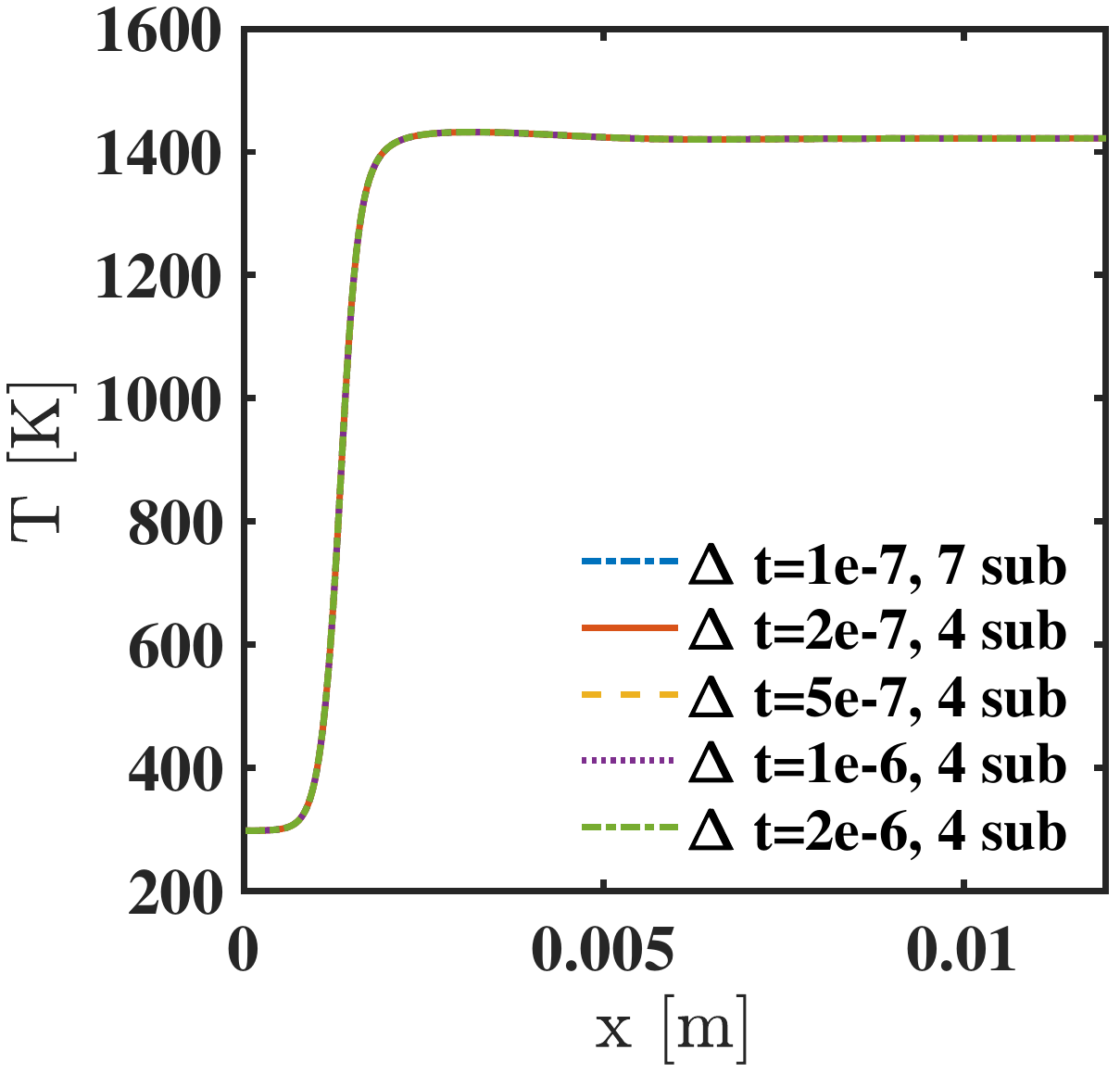}
        \caption{Temperature vs.\ distance}
        \label{Error_mag (a)}
    \end{subfigure}
    \hfill
    \begin{subfigure}{0.49\textwidth}
        \centering
        \includegraphics[width=\textwidth]{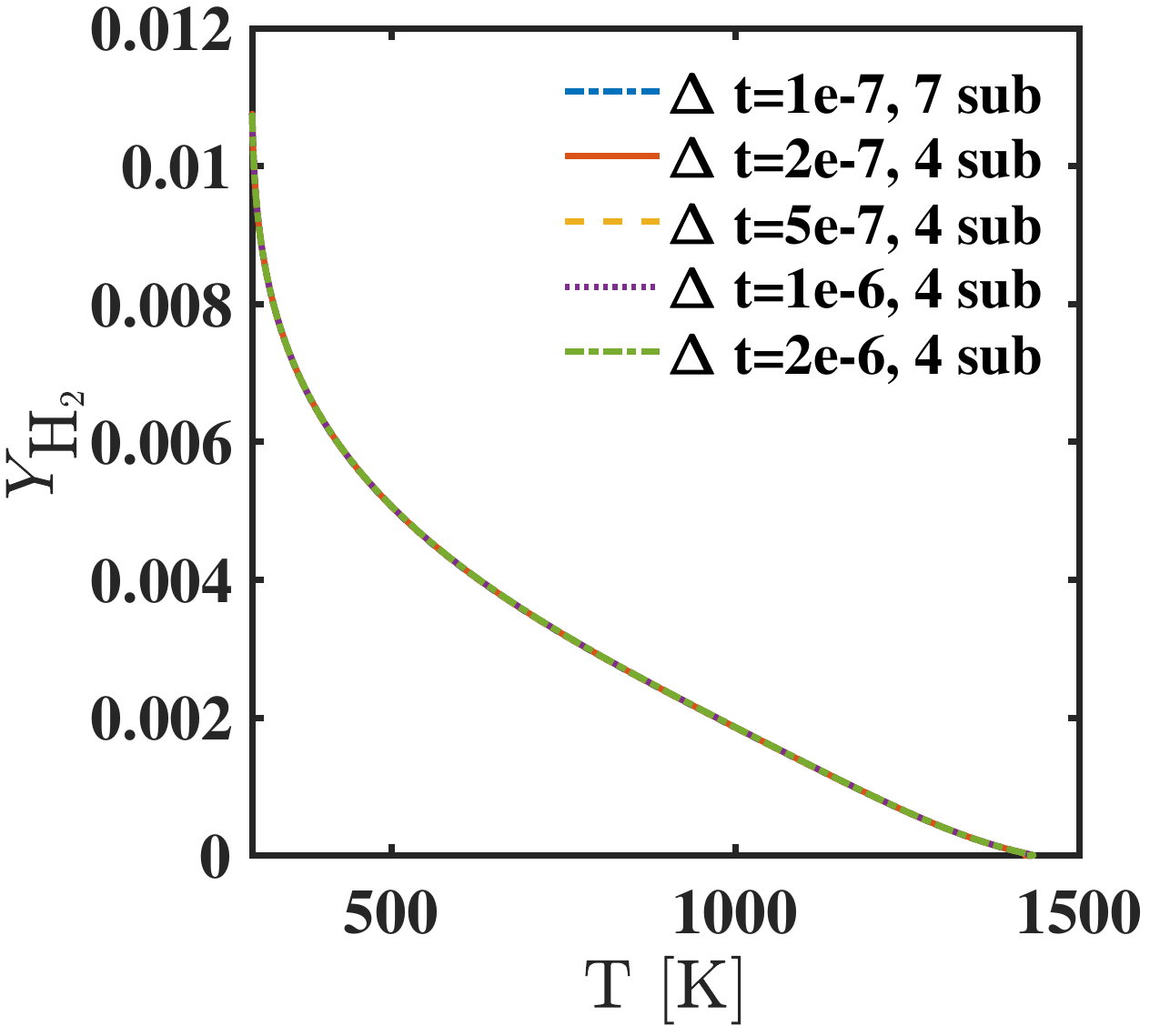}
        \caption{\ce{H2} mass fraction vs.\ temperature}
        \label{Error_mag (b)}
    \end{subfigure}
    \caption{Impact of time-step size and number of sub-iterations on the accuracy of one-dimensional freely propagating flames.}
\end{figure}

To evaluate the of absolute magnitude of error associated with the proposed method, as opposed to the order of accuracy (as time step size approaches zero), Figures~\ref{Error_mag (a)} and \ref{Error_mag (b)} present the temperature as a function of distance and fuel mass fraction as a function of temperature, respectively, for a range of freely propagating flames with several time-step sizes and sub-iterations.
The solutions exhibit negligible error in both temperature and fuel mass fractions for the time-step sizes considered, 
and even when using as few as four sub-iterations; these tests demonstrate the high accuracy and robustness of the proposed method.

% \begin{table}[htb!]
% \scriptsize
%     \caption{Laminar flame speed obtained from simulations with various time steps and number of sub-iterations, Q.}
%     \centering
%     %\begin{tabular}{@{\extracolsep{\fill}}l c c c c c c c c@{}}
%     \begin{tabular}{@{}l c c@{}}
%          \toprule
%          $\Delta t$ [s] & Q & $S_L^o$ [\si{\centi\meter\per\second}])\\
%          \midrule
%          \num{1e-8} & 200 & 17.9648 \\
%          \num{2e-7} & 4 & 17.9648 \\
%          \num{2e-6} & 4 & 17.9649 \\
%          \num{5e-6} & 4 & 17.9649 \\
%          \num{5e-6} & 200 & 17.9651 \\
%          \bottomrule
%     \end{tabular}
%     \label{tab:flame_speed_error}
% \end{table}

\subsection{Three-dimensional assessment of diffusion flux models}
In this section we assess a posteriori the species mass diffusion fluxes in the doubly periodic three-dimensional flames \cite{Burali2016AssessmentFlows,Lapointe2016FuelFlames,Schlup2018ctm}.
Differential diffusion effects cause the instabilities found in lean hydrogen\slash air flames, and at high Karlovitz numbers the turbulence time scales match the order of diffusion time scales.
%The test flame uses a nine-species hydrogen model with 54 reactions of Hong et al.~\cite{Hong2011AnMeasurements,Lam2013AAbsorption,Hong2013OnAbsorption} (forward and backward reactions are counted separately).

To assess the impact of the mixture-averaged and multicomponent mass diffusion models on flame chemistry, we compare a posteriori the turbulent and chemistry statistics.
We allow the flames to develop in a turbulent flow field, and compute the statistics after the transients from the initial flow and scalar fields have advected through the domain. 
As an initial assessment, we calculate the effective turbulent flame propagation speeds:
\begin{equation}
S_{\text{T}} = -\frac{\int_{V} \rho \dot{\omega}_{\ce{H2}} dV}{\rho_{u} Y_{\ce{H2},u} L^2} \;.
\end{equation}

\begin{figure}[htb!] 
  \centering
  \includegraphics[width=0.6\textwidth]{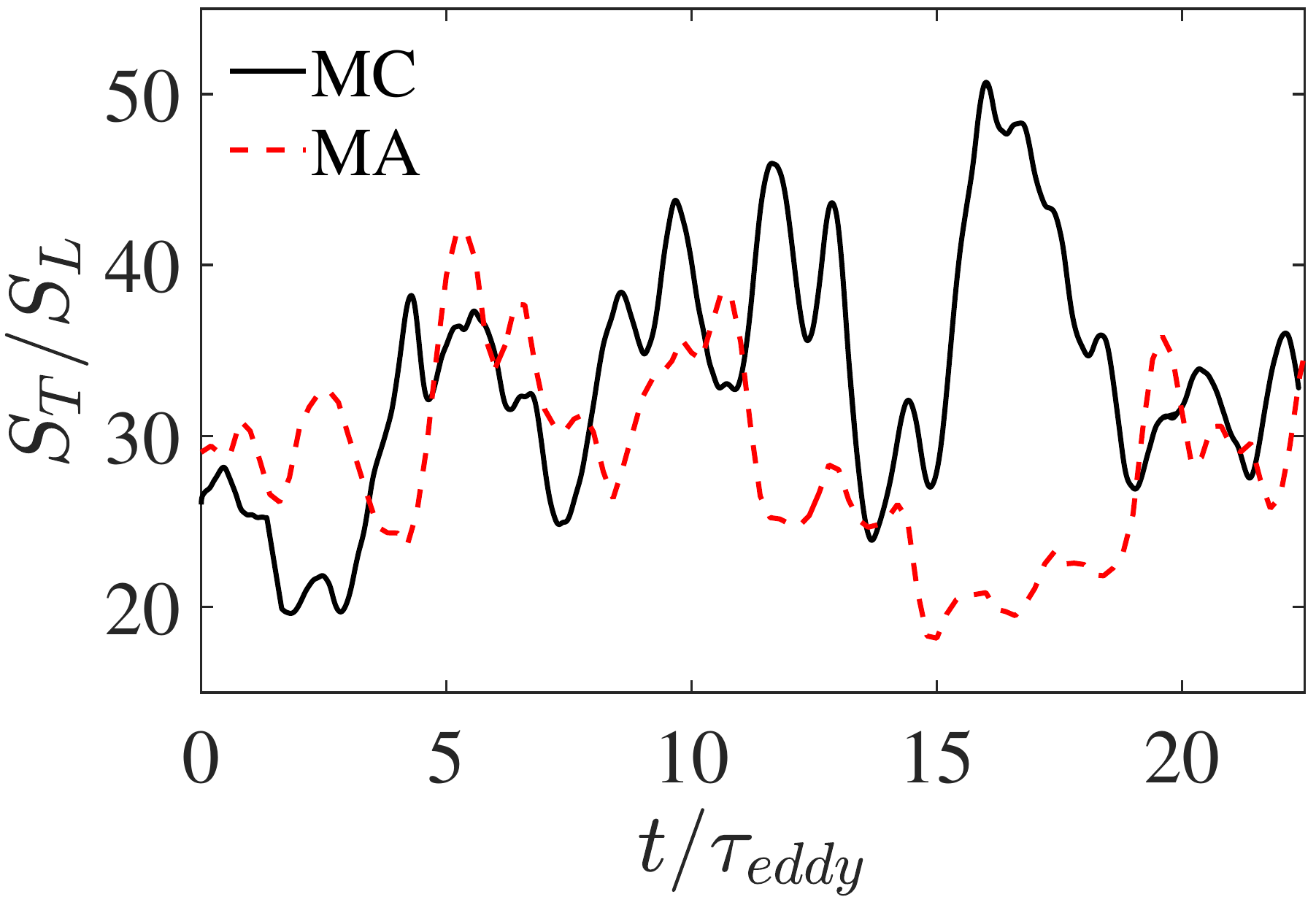}
  \caption{Turbulent flame speed history for three-dimensional, freely propagating, premixed, turbulent hydrogen/air flame with $\phi=0.4$.}
  \label{turbulent_flame_speed}
\end{figure}

Figure~\ref{turbulent_flame_speed} shows the time history of the turbulent flame speed over twenty-two eddy turn-over times ($\tau_{\text{eddy}}$).
The average normalized flames speeds from the mixture-averaged and multicomponent models differ by \SI{15}{\percent}:
$S_{T}^{\text{MA}}/S_{L}=29.6$ and $S_{T}^{\text{MC}}/S_{L}^{0}=34.7$, respectively.
Further study is needed on whether the mixture-averaged diffusion model fully captures the fundamental physics of multicomponent diffusion.

\begin{figure}[htbp]
  \centering
    \begin{subfigure}{0.49\textwidth}
    \centering
    \includegraphics[width=\textwidth]{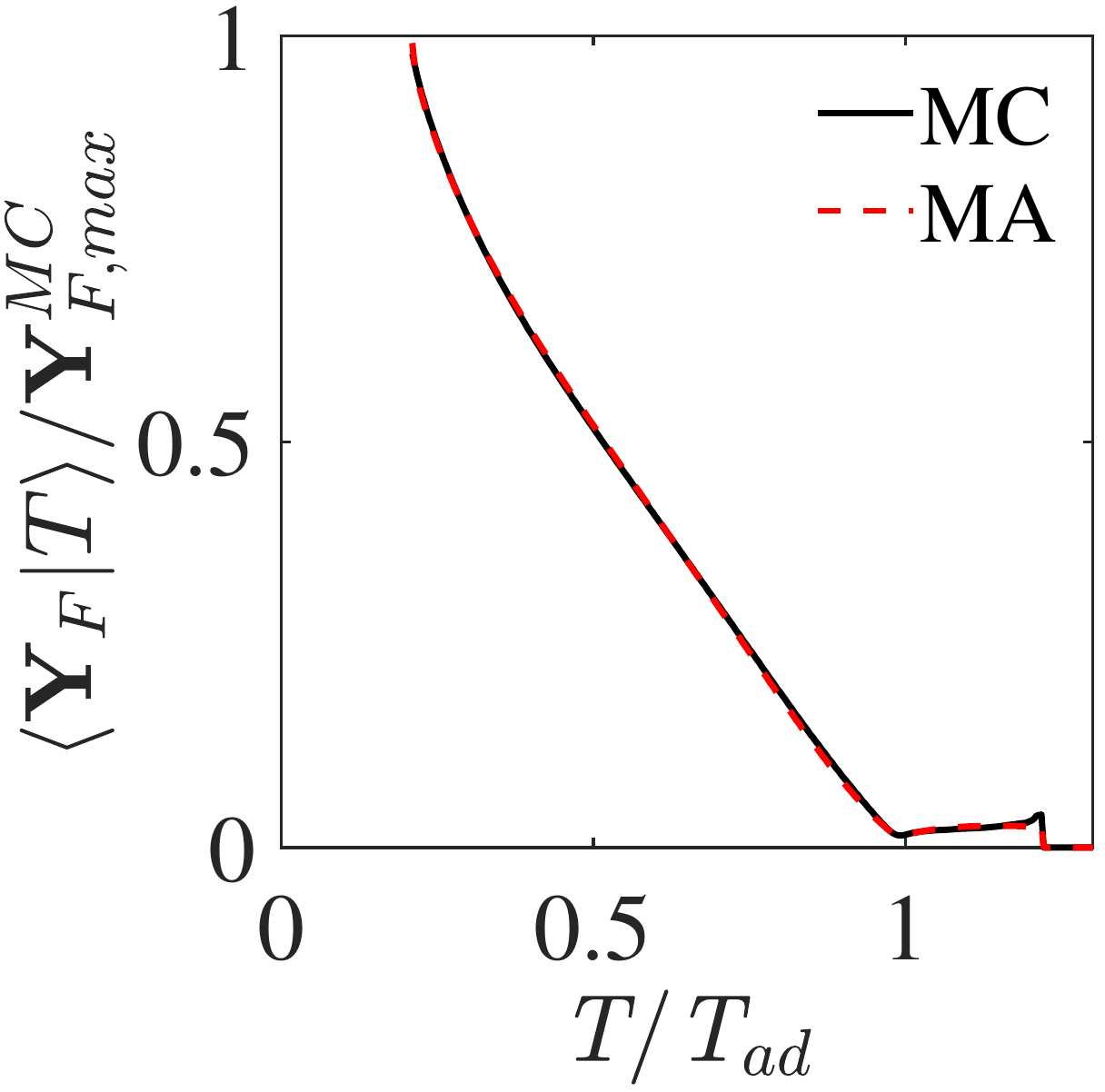}
    \caption{Fuel mass fraction}
    \end{subfigure}
    \hfill
    \begin{subfigure}{0.46\textwidth}
    \centering
    \includegraphics[width=\textwidth]{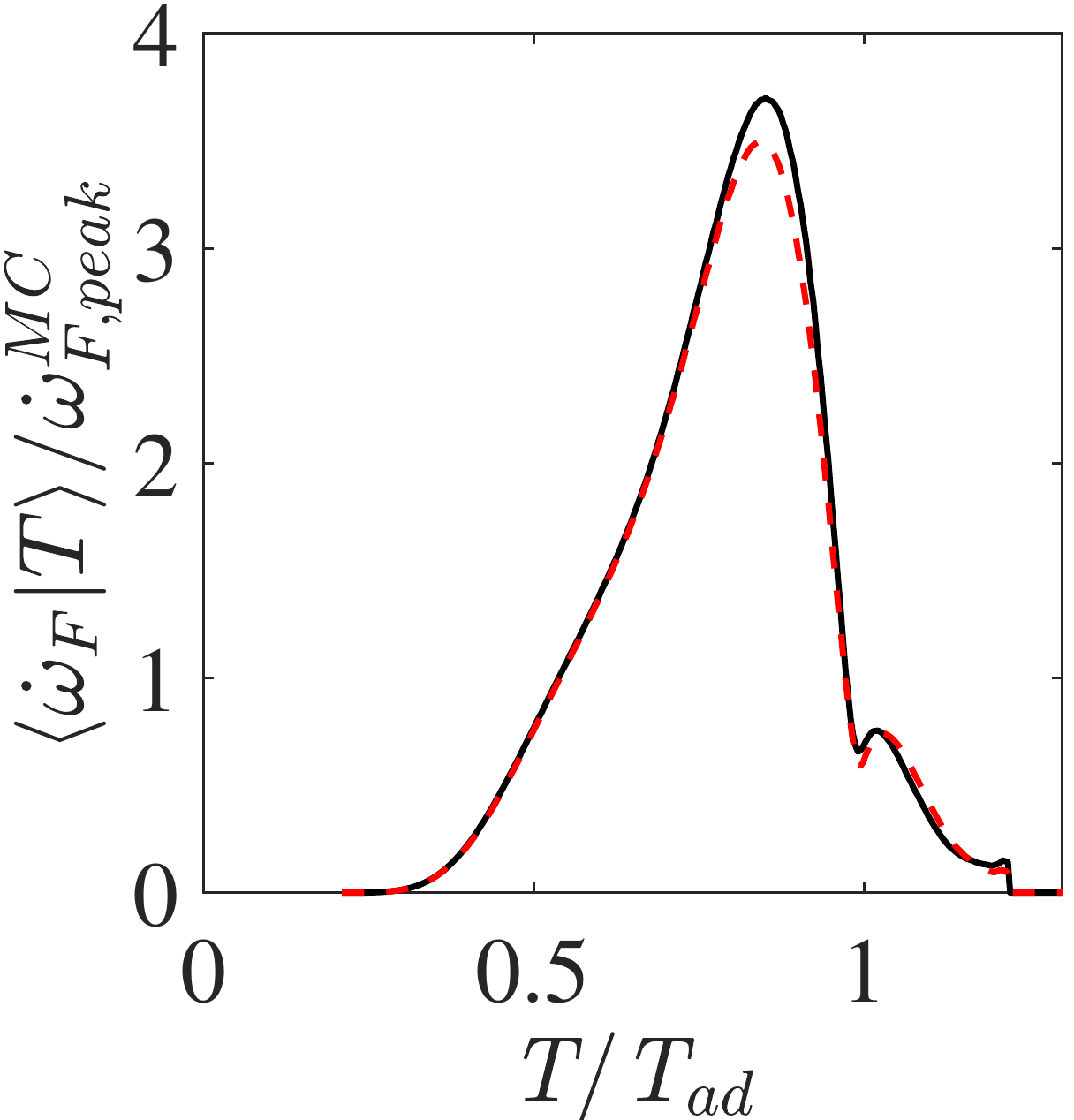}
    \caption{Fuel chemical source term}
    \end{subfigure}
  \caption{Conditional means on temperature for the three-dimensional, freely propagating, premixed, turbulent hydrogen/air flame with $\phi=0.4$.}
  \label{conditional_means}
\end{figure}

To further assess any differences between the mixture-averaged and multicomponent mass diffusion models, Figure~\ref{conditional_means} presents the means of fuel mass fraction and its source term conditioned on temperature for the full time domain.
The differences in the calculated conditional means are small: less than \SI{5.5}{\percent}.
This agreement also extends into super-adiabatic regions for the hydrogen/air flame;
these regions, also called ``hot spots'', result from differential diffusion and have been predicted both in theoretical studies~\cite{Williams:1985} and numerical analyses of lean hydrogen/air mixtures~\cite{Day:2009,AspdenJFM:2011,Aspden:2015}.
However, these small differences in global flame statistics do not explain the \SI{15}{\percent} difference observed in the turbulent flame speeds between the mixture-averaged and multicomponent diffusion models.
These results raise questions on the appropriateness of the mixture-averaged diffusion assumption for direct numerical simulation and warrants further investigation.

\subsection{Computational cost}
This section discusses the relative cost for implementing the full multicomponent mass diffusion to provide context for its use.
The presented timing comparisons examine how the method scales with both number of chemical species and spatial dimension.

\begin{figure}[htbp]
  \centering
  \includegraphics[width=0.6\textwidth]{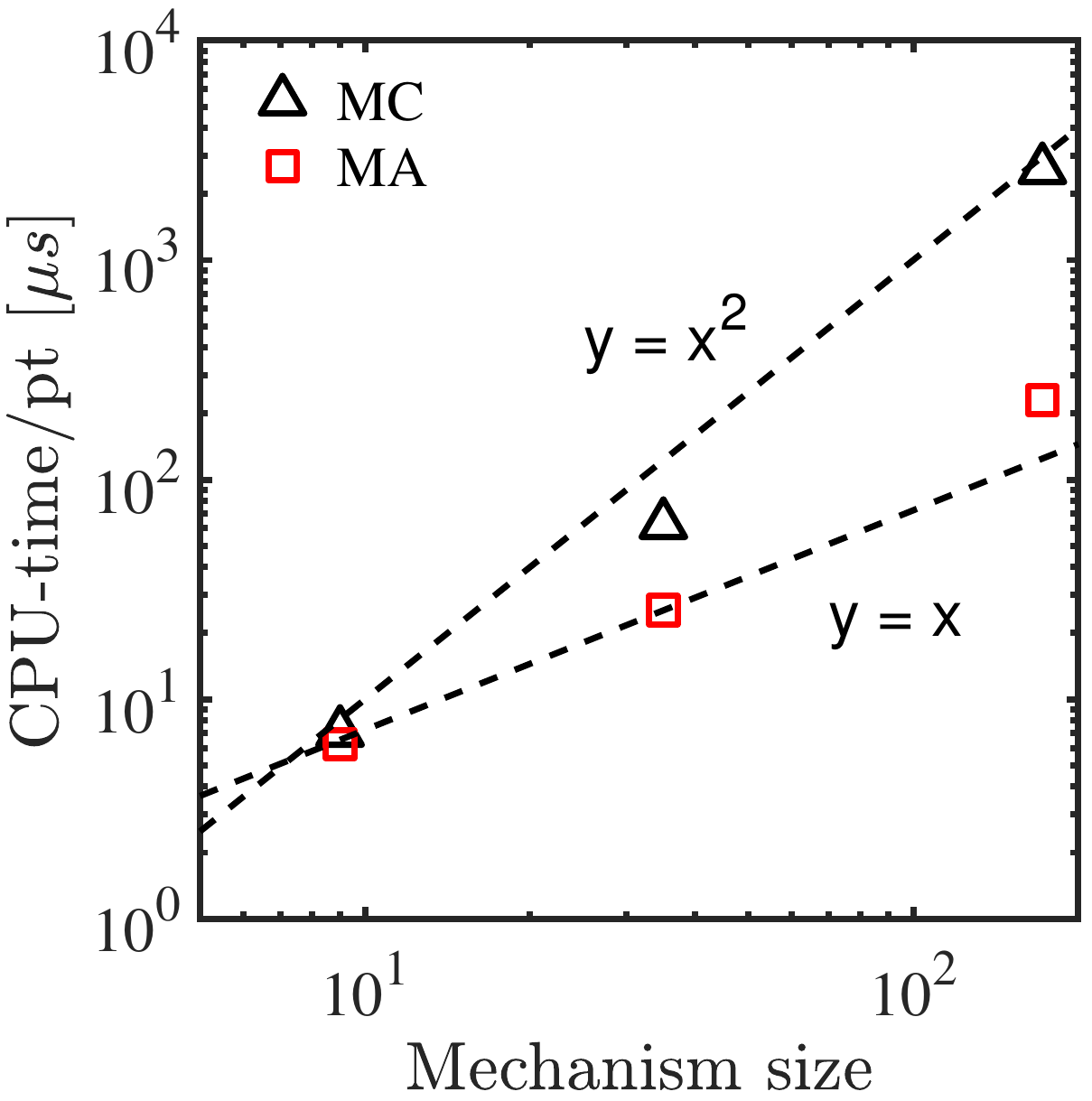}
  \caption{Computational time per grid point for computing diffusion coefficients and diffusion mass fluxes using kinetic models with 9, 35, and 172 species; black dashed lines correspond to linear ($y=x$) and quadratic ($y=x^2$) scaling trends respectively. 
  MC and MA stand for multicomponent and mixture-averaged, respectively.}
  \label{1D_scaling}
\end{figure}

We tested three chemical kinetic models (containing 9 \cite{Burali2016AssessmentFlows}, 35 \cite{Bisetti2012,Savard2015}, and 172 species \cite{Blanquart2007,Blanquart2009ChemicalPrecursors,Narayanaswamy2010ASpecies}) in a one-dimensional flat flame simulation to determine the cost of multicomponent mass diffusion over a wide range of model sizes. %\cite{Hong2011AnMeasurements,Lam2013AAbsorption,Hong2013OnAbsorption}
Figure~\ref{1D_scaling} shows the computational time per grid point for computing the diffusion mass fluxes on a desktop workstation using an Intel Xeon-X5660 CPU with a \SI{2.80}{GHz} clock speed.
The presented timings include calculation of both the diffusion coefficients and mass diffusion fluxes for all aspects of the code.

For the tested chemical kinetic models, the mixture-averaged model scales linearly while the multicomponent model scales quadratically with the number of species.
The multicomponent model is more expensive and does take more time per-point for all three test cases.
For the largest kinetic model (with 172 species) the multicomponent case is noticeably more expensive than the mixture-averaged model.
The increased cost for the multicomponent simulations comes primarily from the CHEMKIN II~\cite{Kee1989Chemkin-II:Kinetics} routine used to determine the ordinary multicomponent diffusion coefficient matrix.

The relevant cost for the proposed method can be split into three primary categories: the costs of calculating the multicomponent diffusion coefficients, calculating the multicomponent diffusion fluxes, and the semi-implicit integration scheme.
Since the proposed method for implementing full multicomponent mass diffusion focuses on efficient low-memory calculation of the diffusion fluxes, rather than the multicomponent diffusion coefficients, the cost of CHEMKIN should be considered independently of the proposed algorithm.
Moreover, the semi-implicit scheme is the same for the mixture-averaged and multicomponent cases, because both cases use the mixture-averaged diffusion coefficient matrix to approximate the Jacobian for the diffusion source terms.
As a result, the two methods have similar implementation and computational expense, with the exception of using CHEMKIN II~\cite{Kee1989Chemkin-II:Kinetics}.

\begin{figure}[htbp]
    \begin{subfigure}{0.49\textwidth}
        \centering
        \includegraphics[width=\textwidth]{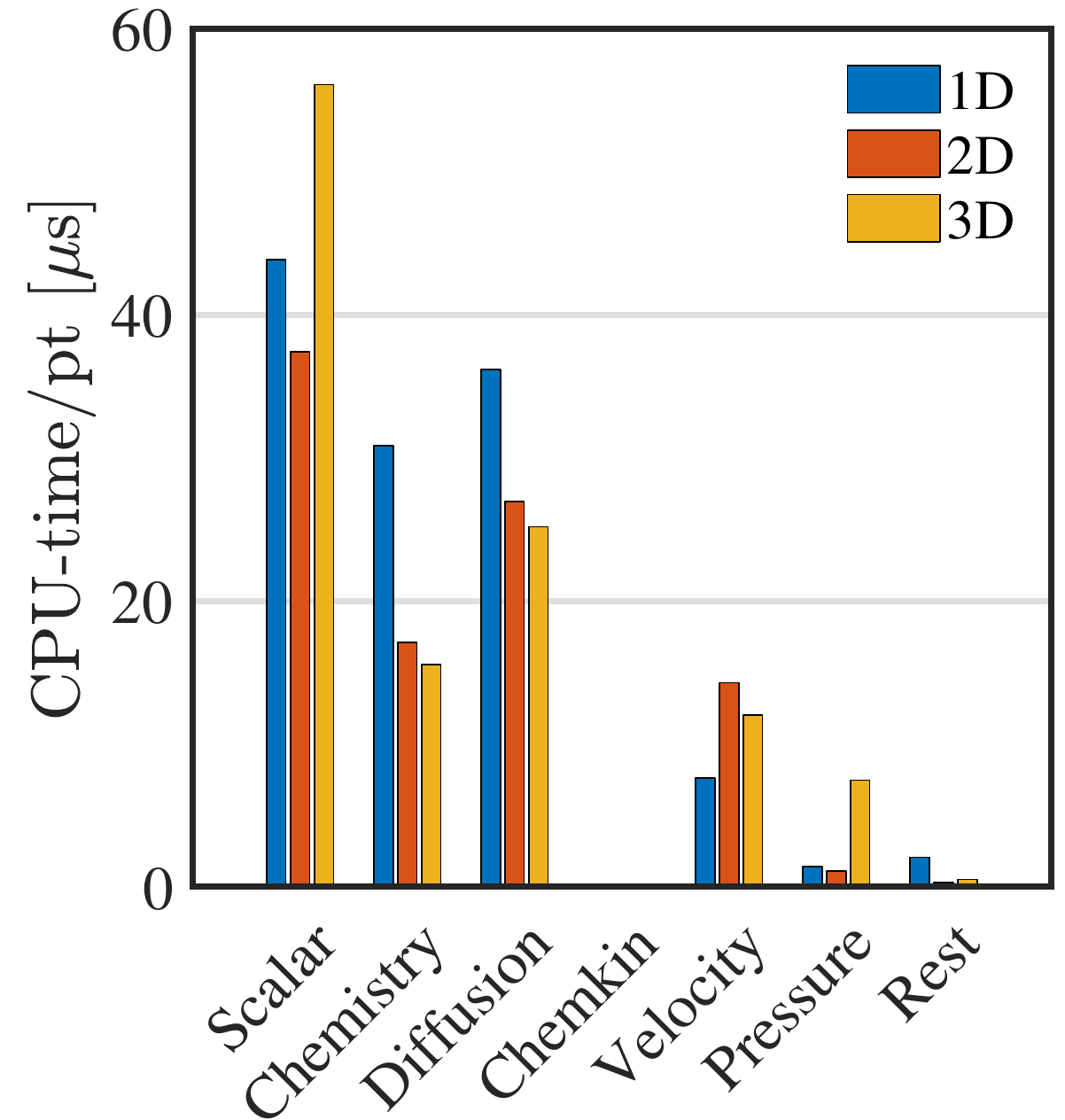}
        \caption{Mixture-averaged model}\label{Scaling (a)}
    \end{subfigure}
    \begin{subfigure}{0.49\textwidth}
        \centering
        \includegraphics[width=\textwidth]{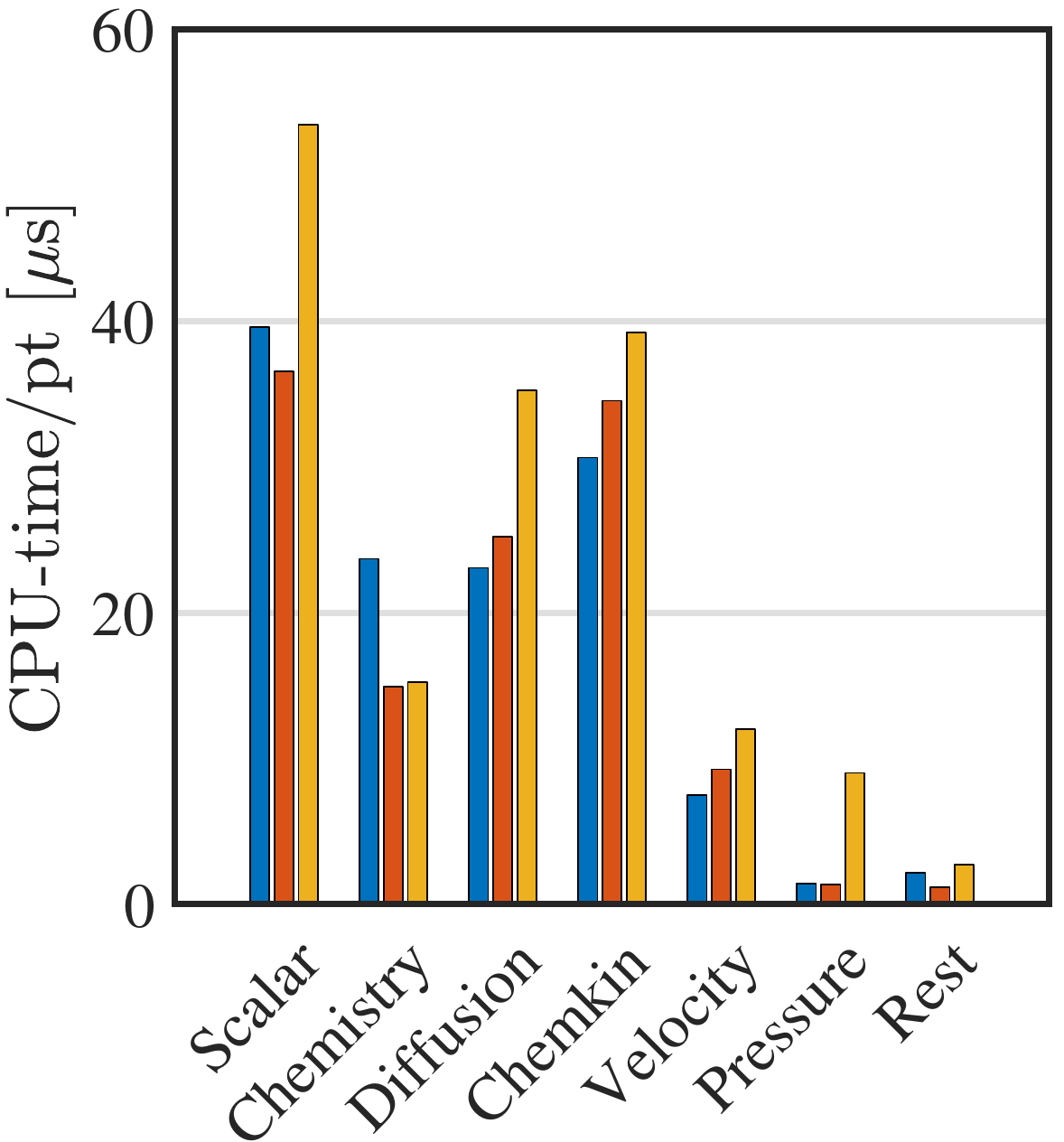}
        \caption{Multicomponent model}\label{Scaling (b)}
    \end{subfigure}
  \caption{Computational time per grid point for each of the three flame configurations: one dimensional (blue), two dimensional (red), and three dimensional (yellow).}
  \label{Scaling}
\end{figure}
\begin{figure}[htbp]
    \centering
    \includegraphics[width=\textwidth]{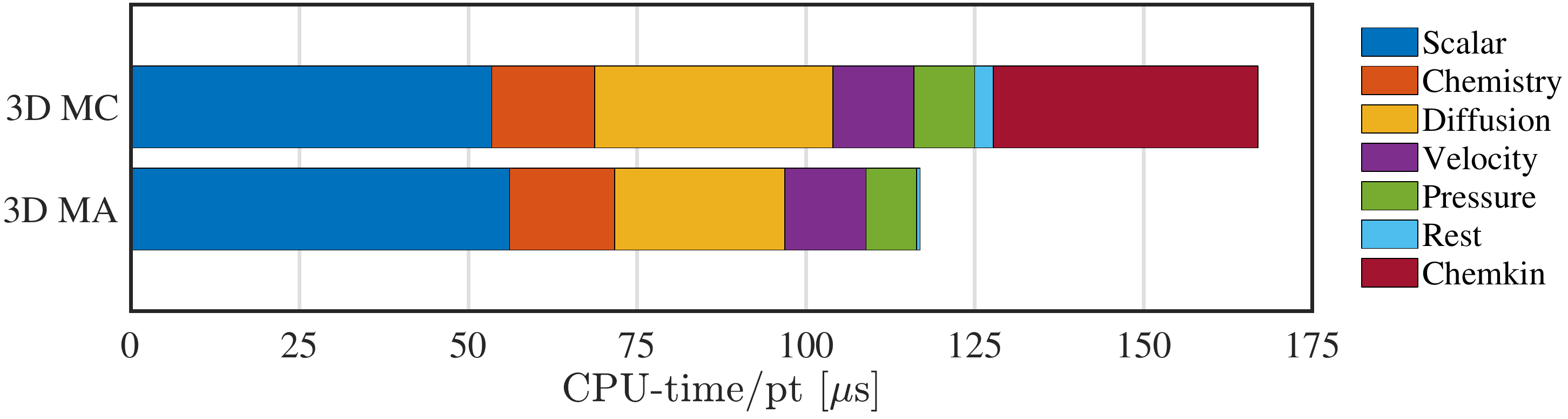}
    \caption{Comparison of numerical costs for the three-dimensional hydrogen flame simulations.}\label{Scaling_comparision}
\end{figure}

To evaluate how the multicomponent model scales with increasing spatial dimension, and evaluate the relative cost of using CHEMKIN II~\cite{Kee1989Chemkin-II:Kinetics}, we acquired timings for one- ($720$ grid points), two- ($1888 \times 472$ grid points), and three-dimensional ($1520 \times 190 \times 190$ grid points) configurations covering the cases presented in this work, with the additional two-dimensional case matching similar timing tests by Schlup et al.~\cite{Schlup2018ctm}.
These timing tests represent an average cost per point and are determined by averaging the timings taken for the 20 time steps, skipping the first and last integrations.
Figure~\ref{Scaling} presents the computational timings for each part of the code for both diffusion models, where ``Scalar'' includes scalar field calculation; ``Diffusion'' includes the flux calculation and $\dma$ calculation for the implicit solver; ``Chemistry'', ``Velocity'', and ``Pressure'' are as named; and ``Rest'' account for any remaining computations.
``Scalar'' includes the semi-implicit solver for integrating the diffusion source terms, while the semi-implicit solvers for chemistry and velocity are included in their named categories.
To facilitate comparison between the two models, Figure~\ref{Scaling_comparision} presents the total computational time per grid point for both three-dimensional hydrogen simulations as a stacked bar chart broken down by each section of code.
We performed these computations on the National Energy Research Scientific Computing Center (NERSC) high-performance computing cluster Cori (Cray XC40)~\cite{He2017}.

While much of the code exhibits a similar cost per grid point, regardless of the dimensionality of the problem, the chemistry is more expensive for the one- and two-dimensional cases.
This cost increase is due to NGA's structure.
NGA was written and optimized for three-dimensional configurations, thus the one- and two-dimensional
cases are artificially more expensive, especially in the chemistry calculations~\cite{Schlup2018ctm}.
In addition, for three dimensions the cost of the pressure solver increases due to using the HYPRE package \cite{Falgout2002}.
The one- and two-dimensional cases both implement an exact FFT-tridiagonal solver, while HYPRE---used for the three-dimensional cases---is iterative and thus more expensive.
Despite the minor increase in cost for the pressure solver in three dimensions, the cost is negligible when considering larger kinetic models (i.e., more than 35 species).

Consistent with Figure~\ref{1D_scaling}, the cost of calculating ``Diffusion'' increases with model complexity; recall that $\dma$ is calculated for both the mixture-averaged and multicomponent solvers.
However, the multicomponent diffusion mass flux calculation represents only \SI{21}{\percent} of the total simulation time for the three-dimensional case.
As expected, the cost of calling CHEMKIN II for the diffusion coefficients is large and accounts for roughly \SI{23}{\percent} of the three-dimensional simulation time.
Interestingly, the cost of diffusion increases only slightly moving from one dimension to two dimensions.
This results from the high efficiency of the dynamic memory-allocation algorithm used to implement this model (see Section~\ref{memory}).
Moreover, the multicomponent diffusion implementation is less expensive than the mixture-averaged model for the one-dimensional case and equivalent in cost for the two-dimensional case.
% The increased cost for three-dimensional multicomponent case is likely due to the minor increase in time needed to resolve the full diffusion Jacobian from its approximation. 
Overall, by reducing memory requirements and optimizing calls to memory, the memory algorithm implemented for the multicomponent model maintains low computational expense.

These results indicate that, for hydrogen-air combustion, the multicomponent model is more expensive than the mixture-averaged model; however, the differences in ``Diffusion'' costs between the two models are due to the use of CHEMKIN II~\cite{Kee1989Chemkin-II:Kinetics}.
Thus, the slowdown could be minimized by implementing a more-efficient package for calculating the mass-diffusion coefficients such as EGLIB~\cite{Ern1995,Ern:1998,Ern:1999}; however, the total cost of computing mass diffusion fluxes remains notable, even for the mixture-averaged case.

\section{Summary and future work}
This article presents an efficient and stable scheme for implementing multicomponent mass diffusion in reacting-flow DNS with minimal memory expense.
The proposed scheme exhibits reasonable computational cost for chemical kinetic models of up to 100 species; this performance could be further improved by implementing a more-efficient method for calculating the multicomponent diffusion coefficient matrix.

The results presented for hydrogen flames suggest that the mixture-averaged mass diffusion model may suffice for DNS of three-dimensional, premixed turbulent flames in the regimes and configurations considered.
However, we observed a \SI{15}{\percent} difference in the turbulent flame speeds between the two models, though the differences in the conditional means of the fuel source term and mass fraction were negligible.
The difference observed in turbulent flame 
speeds raises questions about using the mixture-averaged model in DNS of turbulent reacting flows.
Moreover, the algorithm proposed in this study provides a fast, efficient, method for implementing multicomponent mass diffusion in reacting-flow simulations, which may eliminate the need for the mixture-averaged assumption.
However, despite these results, we do not have sufficient data to draw firm conclusions on the accuracy and appropriateness of mixture-averaged assumptions for all flames (i.e., all fuels, configurations, and regimes).
Additional data are needed from studies of different fuels---namely large hydrocarbons---and kinetic models with more species.
Thus, future work should focus on extending these comparisons to other fuels and flame configurations.

\section*{Acknowledgements}
This material is based upon work supported by the National Science Foundation under Grant No.\ 1314109-DGE. 
This research used resources of the National Energy Research Scientific Computing Center (NERSC), a U.S. Department of Energy Office of Science User Facility operated under Contract No.~DE-AC02-05CH11231.

\section*{References}
\bibliography{mendeley}

\begin{thebibliography}{10}
\expandafter\ifx\csname url\endcsname\relax
  \def\url#1{\texttt{#1}}\fi
\expandafter\ifx\csname urlprefix\endcsname\relax\def\urlprefix{URL }\fi
\expandafter\ifx\csname href\endcsname\relax
  \def\href#1#2{#2} \def\path#1{#1}\fi

\bibitem{Bird1960}
R.~B. Bird, W.~E. Stewart, E.~N. Lightfoot, {Transport Phenomena}, John Wiley
  {\&} Sons, Inc., New York, 1960.

\bibitem{Warnatz1978CalculationFlames}
J.~Warnatz, {Calculation of the Structure of Laminar Flat Flames I: Flame
  Velocity of Freely Propagating Ozone Decomposition Flames}, Berichte der
  Bunsengesellschaft f{\"{u}}r physikalische Chemie 82~(2) (1978) 193--200.
\newblock \href {http://dx.doi.org/10.1002/bbpc.197800010}
  {\path{doi:10.1002/bbpc.197800010}}.

\bibitem{Coltrin1986ADeposition}
M.~E. Coltrin, R.~J. Kee, J.~A. Miller, {A Mathematical Model of Silicon
  Chemical Vapor Deposition}, Journal of The Electrochemical Society 133~(6)
  (1986) 1206.
\newblock \href {http://dx.doi.org/10.1149/1.2108820}
  {\path{doi:10.1149/1.2108820}}.

\bibitem{Lapointe2016FuelFlames}
S.~Lapointe, G.~Blanquart, {Fuel and chemistry effects in high Karlovitz
  premixed turbulent flames}, Combustion and Flame 167 (2016) 294--307.
\newblock \href {http://dx.doi.org/10.1016/j.combustflame.2016.01.035}
  {\path{doi:10.1016/j.combustflame.2016.01.035}}.

\bibitem{Burali2016AssessmentFlows}
N.~Burali, S.~Lapointe, B.~Bobbitt, G.~Blanquart, Y.~Xuan, {Assessment of the
  constant non-unity Lewis number assumption in chemically-reacting flows},
  Combustion Theory and Modelling 20~(4) (2016) 632--657.
\newblock \href {http://dx.doi.org/10.1080/13647830.2016.1164344}
  {\path{doi:10.1080/13647830.2016.1164344}}.

\bibitem{Schlup2018ctm}
J.~Schlup, G.~Blanquart, Validation of a mixture-averaged thermal diffusion
  model for premixed lean hydrogen flames, Combustion Theory and Modelling
  22~(2) (2018) 264--290.
\newblock \href {http://dx.doi.org/10.1080/13647830.2017.1398350}
  {\path{doi:10.1080/13647830.2017.1398350}}.

\bibitem{Coffee:1981}
T.~P. Coffee, J.~M. Heimerl, Transport algorithms for premixed, laminar
  steady-state flames, Combust. Flame 43 (1981) 273--289.
\newblock \href {http://dx.doi.org/10.1016/0010-2180(81)90027-4}
  {\path{doi:10.1016/0010-2180(81)90027-4}}.

\bibitem{Warnatz:1982}
J.~Warnatz, Influence of transport models and boundary conditions on flame
  structure, in: N.~Peters, J.~Warnatz (Eds.), Numerical Methods in Laminar
  Flame Propagation, Vol.~6 of Notes on Numerical Fluid Mechanics,
  Vieweg+Teubner Verlag, Wiesbaden, 1982, pp. 87--111.

\bibitem{Ern:1999}
A.~Ern, V.~Giovangigli, Impact of detailed multicomponent transport on planar
  and counterflow hydrogen/air and methane/air flames, Combustion Science and
  Technology 149~(1-6) (1999) 157--181.
\newblock \href {http://dx.doi.org/10.1080/00102209908952104}
  {\path{doi:10.1080/00102209908952104}}.

\bibitem{Bongers:2003}
H.~Bongers, L.~P.~H. De~Goey, The effect of simplified transport modeling on
  the burning velocity of laminar premixed flames, Combust. Sci. Technol.
  175~(10) (2003) 1915--1928.
\newblock \href {http://dx.doi.org/10.1080/713713111}
  {\path{doi:10.1080/713713111}}.

\bibitem{Xin:2015}
Y.~Xin, W.~Liang, W.~Liu, T.~Lu, C.~K. Law, A reduced multicomponent diffusion
  model, Combust. Flame 162~(1) (2015) 68--74.
\newblock \href {http://dx.doi.org/10.1016/j.combustflame.2014.07.019}
  {\path{doi:10.1016/j.combustflame.2014.07.019}}.

\bibitem{FAGHIH2018175}
M.~Faghih, W.~Han, Z.~Chen, Effects of {Soret} diffusion on premixed flame
  propagation under engine-relevant conditions, Combustion and Flame 194 (2018)
  175--179.
\newblock \href {http://dx.doi.org/10.1016/j.combustflame.2018.04.031}
  {\path{doi:10.1016/j.combustflame.2018.04.031}}.

\bibitem{Charentenay:2002}
J.~De~Charentenay, A.~Ern, Multicomponent transport impact on turbulent
  premixed {H}$_2$/{O}$_2$ flames, Combust. Theor. Model. 6~(3) (2002)
  439--462.
\newblock \href {http://dx.doi.org/10.1088/1364-7830/6/3/304}
  {\path{doi:10.1088/1364-7830/6/3/304}}.

\bibitem{Giovangigli2015MulticomponentFlames}
V.~Giovangigli, {Multicomponent transport in laminar flames}, Proceedings of
  the Combustion Institute 35~(1) (2015) 625--637.
\newblock \href {http://dx.doi.org/10.1016/j.proci.2014.08.011}
  {\path{doi:10.1016/j.proci.2014.08.011}}.

\bibitem{Dworkin2009}
S.~Dworkin, M.~Smooke, V.~Giovangigli, {The impact of detailed multicomponent
  transport and thermal diffusion effects on soot formation in ethylene/air
  flames}, Proceedings of the Combustion Institute 32~(1) (2009) 1165--1172.
\newblock \href {http://dx.doi.org/10.1016/j.proci.2008.05.061}
  {\path{doi:10.1016/j.proci.2008.05.061}}.

\bibitem{LU2009192}
T.~Lu, C.~K. Law, Toward accommodating realistic fuel chemistry in large-scale
  computations, Progress in Energy and Combustion Science 35~(2) (2009)
  192--215.
\newblock \href {http://dx.doi.org/10.1016/j.pecs.2008.10.002}
  {\path{doi:10.1016/j.pecs.2008.10.002}}.

\bibitem{Borghesi2015iAConditions}
G.~Borghesi, J.~Bellan, \textit{A priori} and \textit{a posteriori}
  investigations for developing large eddy simulations of multi-species
  turbulent mixing under high-pressure conditions, Physics of Fluids 27~(3)
  (2015) 035117.
\newblock \href {http://dx.doi.org/10.1063/1.4916284}
  {\path{doi:10.1063/1.4916284}}.

\bibitem{masibellan2013}
E.~Masi, J.~Bellan, K.~G. Harstad, N.~A. Okong’o, Multi-species turbulent
  mixing under supercritical-pressure conditions: modelling, direct numerical
  simulation and analysis revealing species spinodal decomposition, Journal of
  Fluid Mechanics 721 (2013) 578–626.
\newblock \href {http://dx.doi.org/10.1017/jfm.2013.70}
  {\path{doi:10.1017/jfm.2013.70}}.

\bibitem{Ern1995}
A.~Ern, V.~Giovangigli, {Fast and Accurate Multicomponent Transport Property
  Evaluation}, Journal of Computational Physics 120~(1) (1995) 105--116.
\newblock \href {http://dx.doi.org/10.1006/JCPH.1995.1151}
  {\path{doi:10.1006/JCPH.1995.1151}}.

\bibitem{Ern:1998}
A.~Ern, V.~Giovangigli, Thermal diffusion effects in hydrogen-air and
  methane-air flames, Combust. Theor. Model. 2~(4) (1998) 349--372.
\newblock \href {http://dx.doi.org/10.1088/1364-7830/2/4/001}
  {\path{doi:10.1088/1364-7830/2/4/001}}.

\bibitem{Ambikasaran2017AnVelocities}
S.~Ambikasaran, K.~Narayanaswamy, {An accurate, fast, mathematically robust,
  universal, non-iterative algorithm for computing multi-component diffusion
  velocities}, Proceedings of the Combustion Institute 36~(1) (2017) 507--515.
\newblock \href {http://dx.doi.org/10.1016/j.proci.2016.05.055}
  {\path{doi:10.1016/j.proci.2016.05.055}}.

\bibitem{Desjardins2008}
O.~Desjardins, G.~Blanquart, G.~Balarac, H.~Pitsch, {High order conservative
  finite difference scheme for variable density low Mach number turbulent
  flows}, Journal of Computational Physics 227~(15) (2008) 7125--7159.
\newblock \href {http://dx.doi.org/10.1016/j.jcp.2008.03.027}
  {\path{doi:10.1016/j.jcp.2008.03.027}}.

\bibitem{Savard2015AChemistry}
B.~Savard, Y.~Xuan, B.~Bobbitt, G.~Blanquart, A computationally-efficient,
  semi-implicit, iterative method for the time-integration of reacting flows
  with stiff chemistry, Journal of Computational Physics 295 (2015) 740--769.
\newblock \href {http://dx.doi.org/10.1016/j.jcp.2015.04.018}
  {\path{doi:10.1016/j.jcp.2015.04.018}}.

\bibitem{Hirschfelder1954}
J.~O. Hirschfelder, C.~F. Curtiss, R.~B. Bird, {Molecular Theory of Gases and
  Liquids}, Wiley, New York, 1954.

\bibitem{Curtiss1949TransportMixtures}
C.~F. Curtiss, J.~O. Hirschfelder, Transport properties of multicomponent gas
  mixtures, The Journal of Chemical Physics 17~(6) (1949) 550--555.
\newblock \href {http://dx.doi.org/10.1063/1.1747319}
  {\path{doi:10.1063/1.1747319}}.

\bibitem{Grcar2009TheFlames}
J.~F. Grcar, J.~B. Bell, M.~S. Day, The {Soret} effect in naturally
  propagating, premixed, lean, hydrogen-air flames, Proceedings of the
  Combustion Institute 32 (2009) 1173--1180.
\newblock \href {http://dx.doi.org/10.1016/j.proci.2008.06.075}
  {\path{doi:10.1016/j.proci.2008.06.075}}.

\bibitem{Schlup2018cnf}
J.~Schlup, G.~Blanquart, A reduced thermal diffusion model for {H} and {H}$_2$,
  Combustion and Flame 191 (2018) 1--8.
\newblock \href {http://dx.doi.org/10.1016/j.combustflame.2017.12.022}
  {\path{doi:10.1016/j.combustflame.2017.12.022}}.

\bibitem{Kee1989Chemkin-II:Kinetics}
R.~Kee, F.~Rupley, J.~Miller, {Chemkin-II}: A {Fortran} chemical kinetics
  package for the analysis of gas-phase chemical kinetics, Sandia National
  Laboratories Report SAND89-8009 (1989).

\bibitem{Dixon-Lewis1968FlameSystems}
G.~Dixon-Lewis, Flame structure and flame reaction kinetics. {II}. transport
  phenomena in multicomponent systems, Proceedings of the Royal Society of
  London A: Mathematical, Physical and Engineering Sciences 307~(1488) (1968)
  111--135.
\newblock \href {http://dx.doi.org/10.1098/rspa.1968.0178}
  {\path{doi:10.1098/rspa.1968.0178}}.

\bibitem{Xuan2014}
Y.~Xuan, G.~Blanquart, M.~E. Mueller, {Modeling curvature effects in diffusion
  flames using a laminar flamelet model}, Combustion and Flame 161~(5) (2014)
  1294--1309.
\newblock \href {http://dx.doi.org/10.1016/j.combustflame.2013.10.028}
  {\path{doi:10.1016/j.combustflame.2013.10.028}}.

\bibitem{Xuan2015}
Y.~Xuan, G.~Blanquart, Effects of aromatic chemistry-turbulence interactions on
  soot formation in a turbulent non-premixed flame, Proceedings of the
  Combustion Institute 35~(2) (2015) 1911--1919.
\newblock \href {http://dx.doi.org/10.1016/j.proci.2014.06.138}
  {\path{doi:10.1016/j.proci.2014.06.138}}.

\bibitem{Xuan2016}
Y.~Xuan, G.~Blanquart, Two-dimensional flow effects on soot formation in
  laminar premixed flames, Combustion and Flame 166 (2016) 113--124.
\newblock \href {http://dx.doi.org/10.1016/j.combustflame.2016.01.007}
  {\path{doi:10.1016/j.combustflame.2016.01.007}}.

\bibitem{Carroll2013}
P.~L. Carroll, G.~Blanquart, {A proposed modification to Lundgren's physical
  space velocity forcing method for isotropic turbulence}, Physics of Fluids
  25~(10) (2013) 105114.
\newblock \href {http://dx.doi.org/10.1063/1.4826315}
  {\path{doi:10.1063/1.4826315}}.

\bibitem{Verma2014}
S.~Verma, Y.~Xuan, G.~Blanquart, An improved bounded semi-{Lagrangian} scheme
  for the turbulent transport of passive scalars, Journal of Computational
  Physics 272 (2014) 1--22.
\newblock \href {http://dx.doi.org/10.1016/j.jcp.2014.03.062}
  {\path{doi:10.1016/j.jcp.2014.03.062}}.

\bibitem{Mueller2012}
M.~E. Mueller, H.~Pitsch, {LES} model for sooting turbulent nonpremixed flames,
  Combustion and Flame 159~(6) (2012) 2166--2180.
\newblock \href {http://dx.doi.org/10.1016/j.combustflame.2012.02.001}
  {\path{doi:10.1016/j.combustflame.2012.02.001}}.

\bibitem{Bisetti2012}
F.~Bisetti, G.~Blanquart, M.~E. Mueller, H.~Pitsch, {On the formation and early
  evolution of soot in turbulent nonpremixed flames}, Combustion and Flame
  159~(1) (2012) 317--335.
\newblock \href {http://dx.doi.org/10.1016/j.combustflame.2011.05.021}
  {\path{doi:10.1016/j.combustflame.2011.05.021}}.

\bibitem{Savard2015}
B.~Savard, G.~Blanquart, Broken reaction zone and differential diffusion
  effects in high {Karlovitz} \textit{n}-{C}$_7${H}$_{16}$ premixed turbulent
  flames, Combustion and Flame 162~(5) (2015) 2020--2033.
\newblock \href {http://dx.doi.org/10.1016/j.combustflame.2014.12.020}
  {\path{doi:10.1016/j.combustflame.2014.12.020}}.

\bibitem{Herrmann2006}
M.~Herrmann, G.~Blanquart, V.~Raman, {Flux Corrected Finite Volume Scheme for
  Preserving Scalar Boundedness in Reacting Large-Eddy Simulations}, AIAA
  Journal 44~(12) (2006) 2879--2886.
\newblock \href {http://dx.doi.org/10.2514/1.18235}
  {\path{doi:10.2514/1.18235}}.

\bibitem{Pierce2001}
C.~Pierce, {Progress-variable approach for large-eddy simulation of turbulent
  combustion}, Ph.D. thesis, Stanford University (2001).

\bibitem{Shunn2012VerificationSolutions}
L.~Shunn, F.~Ham, P.~Moin, {Verification of variable-density flow solvers using
  manufactured solutions}, Journal of Computational Physics 231~(9) (2012)
  3801--3827.
\newblock \href {http://dx.doi.org/10.1016/j.jcp.2012.01.027}
  {\path{doi:10.1016/j.jcp.2012.01.027}}.

\bibitem{Desjardins2008HighFlows}
O.~Desjardins, G.~Blanquart, G.~Balarac, H.~Pitsch, {High order conservative
  finite difference scheme for variable density low Mach number turbulent
  flows}, Journal of Computational Physics 227~(15) (2008) 7125--7159.
\newblock \href {http://dx.doi.org/10.1016/j.jcp.2008.03.027}
  {\path{doi:10.1016/j.jcp.2008.03.027}}.

\bibitem{Falgout2002}
R.~D. Falgout, U.~M. Yang, {hypre: A Library of High Performance
  Preconditioners}, in: P.~M.~A. Sloot, A.~G. Hoekstra, C.~J.~K. Tan, J.~J.
  Dongarra (Eds.), Computational Science --- ICCS 2002, Springer, Berlin,
  Heidelberg, 2002, pp. 632--641.
\newblock \href {http://dx.doi.org/10.1007/3-540-47789-6_66}
  {\path{doi:10.1007/3-540-47789-6_66}}.

\bibitem{vanderVorst1992}
H.~A. van~der Vorst, Bi-{CGSTAB}: A fast and smoothly converging variant of
  bi-{CG} for the solution of nonsymmetric linear systems, {SIAM} Journal on
  Scientific and Statistical Computing 13~(2) (1992) 631--644.
\newblock \href {http://dx.doi.org/10.1137/0913035}
  {\path{doi:10.1137/0913035}}.

\bibitem{Frigo2005}
M.~Frigo, S.~Johnson, The design and implementation of {FFTW}3, Proceedings of
  the {IEEE} 93~(2) (2005) 216--231.
\newblock \href {http://dx.doi.org/10.1109/jproc.2004.840301}
  {\path{doi:10.1109/jproc.2004.840301}}.

\bibitem{Richardson1911TheDam}
L.~F. Richardson, The approximate arithmetical solution by finite differences
  of physical problems involving differential equations, with an application to
  the stresses in a masonry dam, Philosophical Transactions of the Royal
  Society of London A: Mathematical, Physical and Engineering Sciences
  210~(459-470) (1911) 307--357.
\newblock \href {http://dx.doi.org/10.1098/rsta.1911.0009}
  {\path{doi:10.1098/rsta.1911.0009}}.

\bibitem{Perini2014AMechanisms}
F.~Perini, E.~Galligani, R.~D. Reitz, {A study of direct and Krylov iterative
  sparse solver techniques to approach linear scaling of the integration of
  chemical kinetics with detailed combustion mechanisms}, Combustion and Flame
  161~(5) (2014) 1180--1195.
\newblock \href {http://dx.doi.org/10.1016/j.combustflame.2013.11.017}
  {\path{doi:10.1016/j.combustflame.2013.11.017}}.

\bibitem{Hong2011AnMeasurements}
Z.~Hong, D.~F. Davidson, R.~K. Hanson, An improved {H}$_2$/{O}$_2$ mechanism
  based on recent shock tube/laser absorption measurements, Combustion and
  Flame 158~(4) (2011) 633--644.
\newblock \href {http://dx.doi.org/10.1016/j.combustflame.2010.10.002}
  {\path{doi:10.1016/j.combustflame.2010.10.002}}.

\bibitem{Lam2013AAbsorption}
K.-Y. Lam, D.~F. Davidson, R.~K. Hanson, A shock tube study of {H}$_2$ + {OH}
  $\to$ {H}$_2${O} + {H} using {OH} laser absorption, International Journal of
  Chemical Kinetics 45~(6) (2013) 363--373.
\newblock \href {http://dx.doi.org/10.1002/kin.20771}
  {\path{doi:10.1002/kin.20771}}.

\bibitem{Hong2013OnAbsorption}
Z.~Hong, K.-Y. Lam, R.~Sur, S.~Wang, D.~F. Davidson, R.~K. Hanson, On the rate
  constants of {OH} + {HO}$_2$ and {HO}$_2$ + {HO}$_2$: A comprehensive study
  of {H}$_2${O}$_2$ thermal decomposition using multi-species laser absorption,
  Proceedings of the Combustion Institute 34~(1) (2013) 565--571.
\newblock \href {http://dx.doi.org/10.1016/j.proci.2012.06.108}
  {\path{doi:10.1016/j.proci.2012.06.108}}.

\bibitem{Goodwin2017}
D.~G. Goodwin, H.~Moffat, K., R.~L. Speth,
  \href{https://www.cantera.org}{{Cantera: An Object-oriented Software Toolkit
  for Chemical Kinetics, Thermodynamics, and Transport Processes, Version
  2.3.0}} (2017).
\newblock \href {http://dx.doi.org/10.5281/zenodo.170284}
  {\path{doi:10.5281/zenodo.170284}}.
\newline\urlprefix\url{https://www.cantera.org}

\bibitem{Rosales2005}
C.~Rosales, C.~Meneveau, {Linear forcing in numerical simulations of isotropic
  turbulence: Physical space implementations and convergence properties},
  Physics of Fluids 17~(9) (2005) 095106.
\newblock \href {http://dx.doi.org/10.1063/1.2047568}
  {\path{doi:10.1063/1.2047568}}.

\bibitem{Lapointe:2015}
S.~Lapointe, B.~Savard, G.~Blanquart, Differential diffusion effects,
  distributed burning, and local extinctions in high {Karlovitz} premixed
  flames, Combust. Flame 162~(9) (2015) 3341--3355.
\newblock \href {http://dx.doi.org/10.1016/j.combustflame.2015.06.001}
  {\path{doi:10.1016/j.combustflame.2015.06.001}}.

\bibitem{Williams:1985}
F.~A. Williams, Combustion Theory, Benjamin/Cummings, 1985.

\bibitem{Day:2009}
M.~Day, J.~Bell, P.~Bremer, V.~Pascucci, V.~Beckner, M.~Lijewski, Turbulence
  effects on cellular burning structures in lean premixed hydrogen flames,
  Combust. Flame 156~(5) (2009) 1035--1045.
\newblock \href {http://dx.doi.org/10.1016/j.combustflame.2008.10.029}
  {\path{doi:10.1016/j.combustflame.2008.10.029}}.

\bibitem{AspdenJFM:2011}
A.~J. Aspden, M.~S. Day, J.~B. Bell, Turbulence-flame interactions in lean
  premixed hydrogen: {T}ransition to the distributed burning regime, J. Fluid
  Mech. 680 (2011) 287--320.
\newblock \href {http://dx.doi.org/10.1017/jfm.2011.164}
  {\path{doi:10.1017/jfm.2011.164}}.

\bibitem{Aspden:2015}
A.~Aspden, M.~Day, J.~Bell, Turbulence-chemistry interaction in lean premixed
  hydrogen combustion, Proc. Combust. Inst. 35~(2) (2015) 1321--1329.
\newblock \href {http://dx.doi.org/10.1016/j.proci.2014.08.012}
  {\path{doi:10.1016/j.proci.2014.08.012}}.

\bibitem{Blanquart2007}
G.~Blanquart, H.~Pitsch, {Thermochemical Properties of Polycyclic Aromatic
  Hydrocarbons (PAH) from G3MP2B3 Calculations}, J. Phys. Chem. A 111~(28)
  (2007) 6510--6520.
\newblock \href {http://dx.doi.org/10.1021/JP068579W}
  {\path{doi:10.1021/JP068579W}}.

\bibitem{Blanquart2009ChemicalPrecursors}
G.~Blanquart, P.~Pepiot-Desjardins, H.~Pitsch, {Chemical mechanism for high
  temperature combustion of engine relevant fuels with emphasis on soot
  precursors}, Combustion and Flame 156~(3) (2009) 588--607.
\newblock \href {http://dx.doi.org/10.1016/j.combustflame.2008.12.007}
  {\path{doi:10.1016/j.combustflame.2008.12.007}}.

\bibitem{Narayanaswamy2010ASpecies}
K.~Narayanaswamy, G.~Blanquart, H.~Pitsch, A consistent chemical mechanism for
  oxidation of substituted aromatic species, Combustion and Flame 157~(10)
  (2010) 1879--1898.
\newblock \href {http://dx.doi.org/10.1016/j.combustflame.2010.07.009}
  {\path{doi:10.1016/j.combustflame.2010.07.009}}.

\bibitem{He2017}
Y.~He, B.~Cook, J.~Deslippe, B.~Friesen, R.~Gerber, R.~Hartman-Baker,
  A.~Koniges, T.~Kurth, S.~Leak, W.-S. Yang, Z.~Zhao, E.~Baron, P.~Hauschildt,
  Preparing {NERSC} users for {Cori}, a {Cray} {XC}40 system with {Intel} many
  integrated cores, Concurrency and Computation: Practice and Experience 30~(1)
  (2017) e4291.
\newblock \href {http://dx.doi.org/10.1002/cpe.4291}
  {\path{doi:10.1002/cpe.4291}}.

\bibitem{repropack}
A.~J. Fillo, J.~Schlup, G.~Beardsell, G.~Blanquart, K.~E. Niemeyer, Figures,
  plotting scripts, and data for ``{A} fast, low-cost, and stable memory
  algorithm for implementing multicomponent transport in direct numerical
  simulations'' [dataset], Zenodo (2019).
\newblock \href {http://dx.doi.org/10.5281/zenodo.3519910}
  {\path{doi:10.5281/zenodo.3519910}}.

\bibitem{MCdata}
A.~J. Fillo, J.~Schlup, G.~Beardsell, G.~Blanquart, K.~E. Niemeyer, Direct
  numerical simulation results for turbulent hydrogen/air flame: multicomponent
  diffusion model [dataset], Oregon State University (2019).
\newblock \href {http://dx.doi.org/10.7267/2f75rf80s}
  {\path{doi:10.7267/2f75rf80s}}.

\bibitem{MAdata}
A.~J. Fillo, J.~Schlup, G.~Beardsell, G.~Blanquart, K.~E. Niemeyer, Direct
  numerical simulation results for turbulent hydrogen/air flame:
  mixture-averaged diffusion model [dataset], Oregon State University (2019).
\newblock \href {http://dx.doi.org/10.7267/c247f0072}
  {\path{doi:10.7267/c247f0072}}.

\end{thebibliography}
\bibliographystyle{elsarticle-num}

\appendix
% Fix for missing space between "Appendix" and letter
\renewcommand*{\thesection}{\appendixname~\Alph{section}}

\section{Availability of material}
The figures in this article, as well as the data and plotting scripts necessary to reproduce them, are available 
openly under the CC-BY license~\cite{repropack}.
Furthermore, the full simulation results from NGA are available for the three-dimensional
multicomponent~\cite{MCdata} and mixture-averaged~\cite{MAdata} hydrogen/air flames.

\section{Method verification}\label{method_validation}
\begin{figure}[htb!]
  \centering
  \begin{subfigure}{\textwidth}
        \centering
        \includegraphics{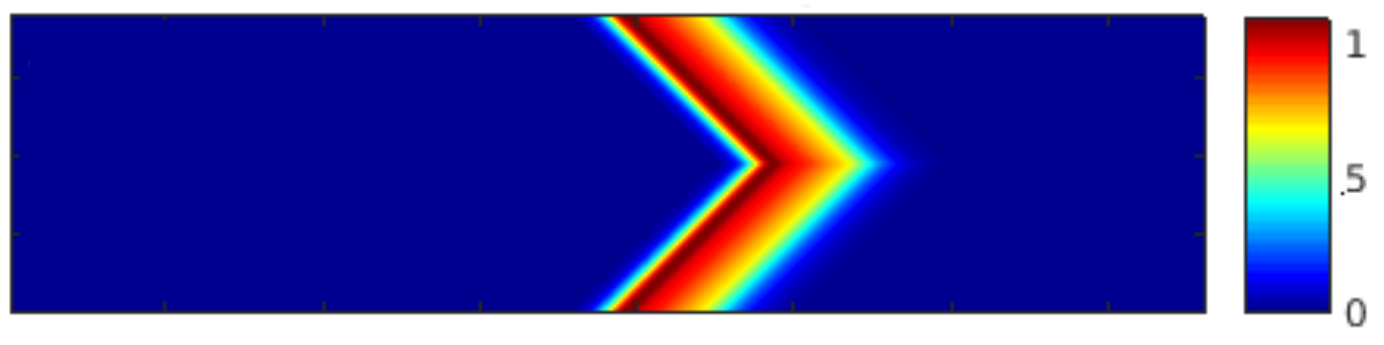}
        \caption{Input species profile}
        \label{input_test_profile}
    \end{subfigure}
    \\
    \begin{subfigure}{\textwidth}
        \centering
        \includegraphics{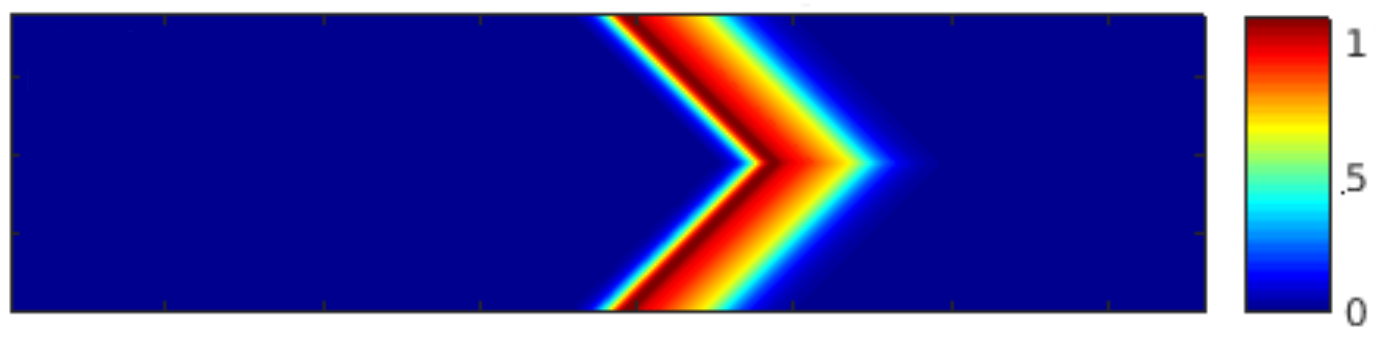}
        \caption{$x$-component of mass flux}
    \end{subfigure}
    \\
    \begin{subfigure}{\textwidth}
        \centering
        \includegraphics{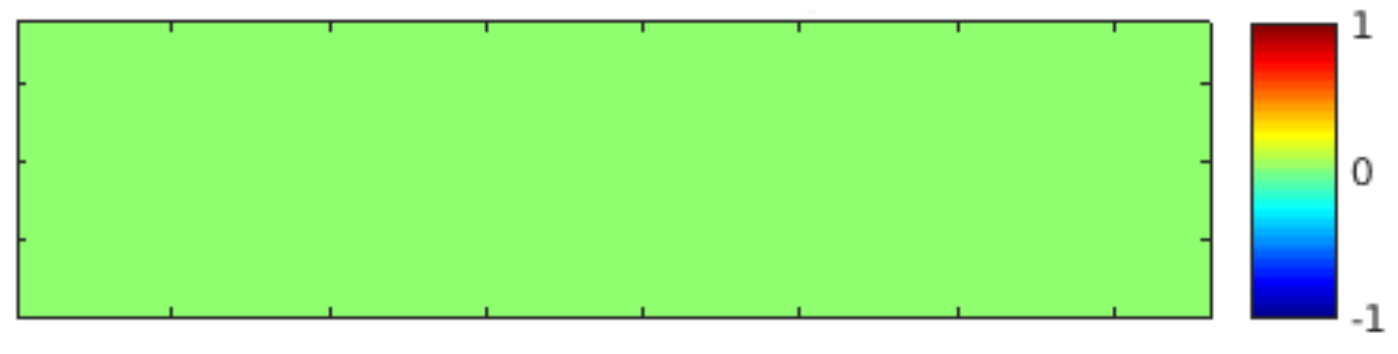}
        \caption{$y$-component of mass flux}
    \end{subfigure}
    \\
    \begin{subfigure}{\textwidth}
        \centering
        \includegraphics{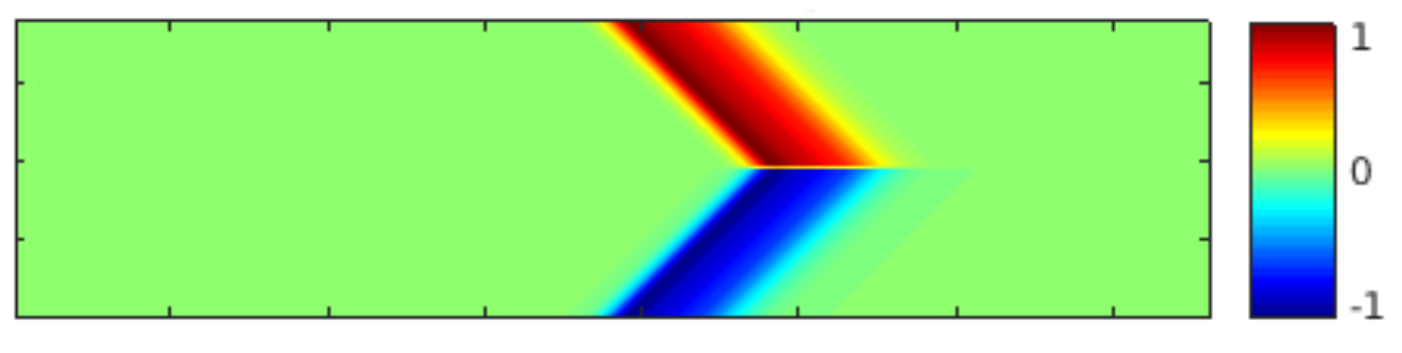}
        \caption{$z$-component of mass flux}
    \end{subfigure}
  \caption{Normalized flux vectors resulting from an artificial species profile after one full iteration of semi-implicit multicomponent diffusion calculation.}
  \label{output_test_profiles}
\end{figure}

To verify the method implementation, we generated an artificial species profile where the direction and relative magnitudes of the flux vectors could be predicted a priori to remain independent of any differential diffusion effects that may exist in a physical system.
Specifically, we created a two-dimensional V-shaped species profile with a central angle of \SI{45}{\degree} and projected it into three dimensions as shown in Figure~\ref{input_test_profile}.

% \begin{figure}[htb]
%   \centering
%   \includegraphics[width=0.9\textwidth]{xz_plane_vFlame.pdf}
%   \caption{Normalized artificial input species profile for method verification; slice from $x$-$z$-plane.}
%   \label{input_test_profile}
% \end{figure}

Such a profile results in flux vectors that are constant in the $y$-direction, are of equal magnitude and opposite sign in the $z$-direction reflected over the $x$-$y$-plane, and vary in magnitude but remain constant in sign matching the initial input profile in the $x$-direction.
These predictions should be consistent independent of chemical species or other scalar value for the artificial input profile.
We ran the algorithm for one ``complete'' set of sub-iterations to convergence and normalized the resulting diffusion flux vectors to ensure the relative magnitudes and direction were consistent with our expectations.

Figure~\ref{output_test_profiles} shows the results of this artificial test case.
The resulting normalized flux vectors agree with expectation and have equal magnitudes in the $x$- and $z$-directions corresponding to the \SI{45}{\degree} artificial flame angle. 
This result indicates proper functionality of the proposed method.

\end{document}